\documentclass[conference]{IEEEtran}
\IEEEoverridecommandlockouts
\pdfoutput=1
\usepackage{cite}
\usepackage{amsmath,amssymb,amsfonts}
\usepackage{algorithmic}
\usepackage{textcomp}
\usepackage{xcolor}
\usepackage{multirow}

\makeatletter
\let\MYcaption\@makecaption
\makeatother
\usepackage{subcaption}
\usepackage{graphicx}

\allowdisplaybreaks[1]
\newcommand{\etal}[0]{\textit{et \allowbreak al.}}

\newcommand\Fig[1] {Fig.~\ref{#1}}
\newcommand\Figs[1] {Figs.~\ref{#1}}

\newcommand{\argmax}{\operatornamewithlimits{argmax}}

\newcommand{\Dists}{\mathbb{D}} 

\newcommand{\expect}{\operatornamewithlimits{\mathbb{E}}}

\newcommand{\nats}{\mathbb{N}}

\newcommand{\realsnng}{\mathbb{R}_{\ge0}}

\newcommand{\calx}{\mathcal{X}}
\newcommand{\caly}{\mathcal{Y}}

\newcommand{\Atk}{\mathit{A}}

\newcommand{\Lin}{\mathsf{in}}
\newcommand{\Lout}{\mathsf{out}}

\newcommand{\ME}{\mathit{mpe}}

\newcommand{\Dtrain}[0]{D^{\sf train}}
\newcommand{\Dtest}[0]{D^{\sf test}}

\newcommand{\calyS}{\mathcal{Y}_{\sf s}}
\newcommand{\calyT}{\mathcal{Y}_{\sf t}}
\newcommand{\PS}{\mathit{P}_{\sf s}}
\newcommand{\PT}{\mathit{P}_{\sf t}}
\newcommand{\fS}{\mathit{f}_{\sf s}}
\newcommand{\fT}{\mathit{f}_{\sf t}}
\newcommand{\tS}{\mathit{t}_{\sf \,s}}
\newcommand{\tT}{\mathit{t}_{\sf \,t}}
\newcommand{\gS}{\mathit{g}_{\sf s}}
\newcommand{\gT}{\mathit{g}_{\sf t}}
\newcommand{\hS}{\mathit{h}_{\sf s}}
\newcommand{\hT}{\mathit{h}_{\sf t}}
\newcommand{\DStrain}[0]{D_{\sf s}^{\sf train}}
\newcommand{\DStest}[0]{D_{\sf s}^{\sf test}}
\newcommand{\DTtrain}[0]{D_{\sf t}^{\sf train}}
\newcommand{\DTtest}[0]{D_{\sf t}^{\sf test}}

\newcommand{\cSH}[0]{\mathit{c}_{\sf shadow}}
\newcommand{\fSH}[0]{\mathit{f}_{\sf shadow}}
\newcommand{\gSH}{\mathit{g}_{\sf shadow}}
\newcommand{\hSH}{\mathit{h}_{\sf shadow}}
\newcommand{\DSHtrain}[0]{D_{\sf shadow}^{\sf train}}
\newcommand{\DSHtest}[0]{D_{\sf shadow}^{\sf test}}
\newcommand{\DSHItrain}[0]{D_{\sf shadow}^{{\sf train},i}}
\newcommand{\DSHItest}[0]{D_{\sf shadow}^{{\sf test},i}}

\newcommand{\fSHS}[0]{\mathit{f}_{\sf s,shadow}}
\newcommand{\gSHS}{\mathit{g}_{\sf s,shadow}}

\newcommand{\fSHT}[0]{\mathit{f}_{\sf t,shadow}}
\newcommand{\gSHT}{\mathit{g}_{\sf t,shadow}}

\newcommand{\DAtrain}[0]{D_{\sf attack}^{\sf train}}

\newcommand{\cstart}[0]{C_{\sf small}}
\newcommand{\cbig}[0]{C_{\sf big}}

\newcommand{\BaselineMIA}{\textsf{BaseMIA}}

\newcommand{\TransMIA}{\textsf{TransMIA}}
\newcommand{\TransMIAL}{\textsf{TransMIA}$_{\sf L}$}
\newcommand{\TransMIAE}{\textsf{TransMIA}$_{\sf E}$}
\newcommand{\TransMIADNN}{\textsf{TransMIA}$_{\sf L(DNN)}$}
\newcommand{\TransMIASVM}{\textsf{TransMIA}$_{\sf L(SVM)}$}

\newcommand{\MIAtL}{\textsf{MIA}$_{\sf L}^{\fT}$}

\newcommand{\MIAtDNN}{\textsf{MIA}$_{\sf L(DNN)}^{\fT}$}

\newif\ifcommentson\commentsonfalse

\newif\ifcamerareadyon\camerareadyonfalse
\ifcamerareadyon
\newcommand{\cameraready}[1]{#1}
\else
\newcommand{\cameraready}[1]{}
\fi

\newif\ifconferenceon\conferenceontrue
\ifconferenceon
\newcommand{\arxiv}[1]{}
\newcommand{\conference}[1]{#1}
\newcommand{\conferenceShort}[1]{}
\else
\newcommand{\arxiv}[1]{#1}
\newcommand{\conference}[1]{}
\newcommand{\conferenceShort}[1]{}
\fi
\ifcommentson
\newcommand{\commentsize}[0]{.90\textwidth}
\newcommand{\commentSH}[1]{\begin{center} \parbox{\commentsize}{\textbf{\textcolor{black}{Comment S.}} \textcolor{red}{#1} }\end{center}}
\newcommand{\commentYK}[1]{\begin{center} \parbox{\commentsize}{\textbf{\textcolor{black}{Comment Y.}} \textcolor{red}{#1} }\end{center}}
\newcommand{\commentTM}[1]{\begin{center} \parbox{\commentsize}{\textbf{\textcolor{black}{Comment T.}} \textcolor{red}{#1 }}\end{center}}
\newcommand{\replySH}[1]{\begin{center} \parbox{\commentsize}{\textbf{Reply S.} \textcolor{blue}{#1} }\end{center}}
\newcommand{\replyYK}[1]{\begin{center} \parbox{\commentsize}{\textbf{Reply Y.} \textcolor{blue}{#1} }\end{center}}
\newcommand{\replyTM}[1]{\begin{center} \parbox{\commentsize}{\textbf{Reply T.} \textcolor{blue}{#1} }\end{center}}
\newcommand{\commentS}[1]{\marginpar{\footnotesize \color{red} {\bf S:} \textsf{\scriptsize #1}}}
\newcommand{\commentY}[1]{\marginpar{\footnotesize \color{red} {\bf Y:} \textsf{\scriptsize #1}}}
\newcommand{\commentT}[1]{\marginpar{\footnotesize \color{red} {\bf T:} \textsf{\scriptsize #1}}}
\newcommand{\replyS}[1]{\marginpar{\footnotesize \color{red} {\bf S:} \textsf{\scriptsize #1}}}
\newcommand{\replyY}[1]{\marginpar{\footnotesize \color{red} {\bf Y:} \textsf{\scriptsize #1}}}
\newcommand{\replyT}[1]{\marginpar{\footnotesize \color{red} {\bf T:} \textsf{\scriptsize #1}}}
\else
\newcommand{\commentSH}[1]{}
\newcommand{\commentYK}[1]{}
\newcommand{\commentTM}[1]{}
\newcommand{\replySH}[1]{}
\newcommand{\replyYK}[1]{}
\newcommand{\replyTM}[1]{}
\newcommand{\commentS}[1]{}
\newcommand{\commentY}[1]{}
\newcommand{\commentT}[1]{}
\newcommand{\replyS}[1]{}
\newcommand{\replyY}[1]{}
\newcommand{\replyT}[1]{}
\fi

\newcommand{\colorR}[1]{\textcolor{red}{#1}}

\newcommand{\pagelimitmarker}[1]{~\\ {\colorR{\ifthenelse{\thepage>#1}{\Huge Exceeding the page limit}{\huge Within the page limit}}}~\\ {\huge{\colorR{~~Page Limit\,\,\,\,\, = #1}}}~\\ {\huge{\colorR{~~Current Page = $\thepage$}}}}

\begin{document}

\title{TransMIA: Membership Inference Attacks Using Transfer Shadow Training}

\author{\IEEEauthorblockN{Seira Hidano}
\IEEEauthorblockA{
\textit{KDDI Research, Inc.}, Saitama, Japan \\
se-hidano at kddi-research.jp}
\and
\IEEEauthorblockN{Takao Murakami}
\IEEEauthorblockA{
\textit{AIST}, Tokyo, Japan}
\and
\IEEEauthorblockN{Yusuke Kawamoto}
\IEEEauthorblockA{
\textit{AIST}, Tokyo, Japan}
}

\maketitle

\begin{abstract}
Transfer learning has been widely studied and gained increasing popularity to improve the accuracy of machine learning models by transferring some knowledge acquired in different training.
However, no prior work has pointed out that transfer learning can strengthen privacy attacks on machine learning models.
In this paper, we propose \textit{TransMIA (Transfer learning-based Membership Inference Attacks)}, which use transfer learning to perform membership inference attacks on the source model when the adversary is able to access the parameters of the transferred model. 
In particular, we propose a \textit{transfer shadow training} technique, 
where an adversary employs the parameters of the transferred model to construct shadow models, to significantly improve the performance of membership inference when a limited amount of shadow training data is available to the adversary.
We evaluate our attacks using two real datasets, and show that our attacks outperform the state-of-the-art that does not use our transfer shadow training technique. 
We also compare four combinations of the learning-based/entropy-based approach and the fine-tuning/freezing approach, all of which employ our transfer shadow training technique.
Then we examine the performance of these four approaches based on the distributions of confidence values, and discuss possible countermeasures against our attacks.
\end{abstract}


\section{Introduction}
\label{sec:intro}
A number of recent machine learning (ML) systems 
are built upon a large amount of training data. 
In many practical applications, however, we may not have sufficient training data due to high expense or difficulty in data acquisition. 
\emph{Transfer learning} \cite{Caruana:94:NIPS,Bengio:11:JMLR,Bengio:12:JMLR,Pan_TKDE10,Weiss:16:JBD,Tan_arXiv18} has been widely studied to improve the accuracy of machine learning systems by 
incorporating some knowledge gained through solving a different problem.
For example, we may transfer a part of an ML model 
trained by a data-rich organization 
(e.g., big company, large hospital) 
to 
another ML model
in a different organization 
(e.g., startup company, small hospital) that has only
a small amount of data 
and wants 
to significantly improve the accuracy of the latter model. 
We refer to the former model as a \emph{source model}, and the latter as a \emph{target model}. 

In many practical ML systems, training data involve sensitive personal data; e.g., face images, medical data. 
A large dataset in the source data-rich organization can also include some sensitive personal data; e.g., face images in a large-scale image dataset for object recognition.
When a source model is trained using sensitive personal data,
transfer learning can raise a serious privacy concern. 
In particular, 
General Data Protection Regulation (GDPR) \cite{GDPR} requires an adequate level of protection when an organization transfers personal data to another organization. 
Therefore, protecting privacy of users in the source model is an important issue in transfer learning across organizations. 

Privacy in transfer learning has been recently studied in the literature. 
For example, privacy-preserving transfer learning algorithms have been proposed in \cite{Gao_BigData19,Guo_arXiv18,Papernot_ICLR17,Wang_ECMLPKDD18,Yao_IJCAI19}. 
A privacy attack against transfer learning 
has also been studied in a recent paper \cite{Zou_arXiv20}. 
Specifically, the authors in \cite{Zou_arXiv20} studied a risk of \emph{membership inference} \cite{Shokri_SP17}, 
in which an adversary infers whether or not personal data of a particular user is used for training a source model. 
They showed by experiments that the adversary who has black-box access to a target model (i.e., who can 
obtain
the target model's output upon an arbitrary input) 
infers membership information in the target model with high accuracy, but cannot infer membership information in the source model 
in the black-box access setting. 

However, it remains open whether the adversary can infer membership information of the source model in the white-box access setting, in which the adversary can access the architecture and parameters of the transferred model (e.g., a part of the source model). 
The privacy risk analysis in the white-box access setting is important to understand 
the ``adequate'' level of protection in transfer learning across organizations. 
The analysis in the white-box setting is also essential in that all the parameters of the transferred model may be leaked by the target organization via illegal access or internal fraud. 
This issue cannot be overlooked, since the number of data breaches is rapidly increasing in recent years~\cite{data_breach1,data_breach2}. 

In this paper, we focus on \emph{deep learning}, one of the most popular approaches in machine learning, and study the risk of membership inference against transfer learning systems in the white-box access setting. 
Specifically, we focus on \emph{network-based deep transfer learning} \cite{Bengio:11:JMLR,Bengio:12:JMLR,Tan_arXiv18}, 
which trains a deep neural network in the source domain (i.e., source model) and reuses a part of the neural network to construct another neural network in the target domain (i.e., target model). 
Then we 
model and analyze an adversary who 
has 
white-box access to the transferred part of the source model.

In this analysis, we assume that 
the source organization does not reveal all parameters of the whole source model to general consumers, since otherwise membership inference attacks would be performed efficiently \cite{Nasr_SP19}.
Then the adversary does not have white black-box access to the whole source model.
Instead, we assume that the adversary has white-box access only to the transferred part of the source model
to analyze 
the \emph{privacy risk in model transfer across organizations}; e.g., 
the privacy risk when 
the target organization, which obtains the transferred model by signing a contract with the source organization, leaks the transferred model 
by illegal access or internal fraud. 
In this setting, the adversary 
obtains the source model's output 
by querying
arbitrary input, and
infers whether or not personal data of a particular user have been used to train the source model. 

One might think that 
if the adversary 
has 
the white-box access to the transferred part of the source model,
then the adversary can easily perform the membership inference attack without 
the black-box access to the whole source model. 
However, 
this is not the case. 
Specifically, Nasr \etal{} \cite{Nasr_SP19} 
showed that 
a white-box membership inference attack that uses the outputs of individual layers 
does not provide better accuracy than the black-box attack. 
Then they proposed a white-box attack using the gradient of the prediction loss with regard to all parameters, and showed that the proposed attack outperforms the black-box attack. 
However, this attack cannot be performed in our setting where the adversary does not have a white-box access to the whole source model, because the adversary cannot calculate the gradient by the back-propagation algorithm in this case. 
For this reason, we
explore a new membership inference algorithm under the assumption of the white-box access to the transferred model and the black-box access to the whole source model. 

Under this assumption, 
we show that \emph{the adversary can also transfer the knowledge} 
in the same way as the target organization to significantly improve the accuracy of membership inference attacks. 
Specifically, we propose a new technique called \emph{transfer shadow training}, which transfers the parameters of the transferred model to a shadow model \cite{Shokri_SP17} that imitates the behavior of the source model and for which the adversary knows the training dataset. 
This technique is helpful for the adversary especially when the amount of shadow training data is small. 
For example, the adversary 
may not have sufficient shadow training data due to high expense or difficulty in acquiring 
personal data 
(e.g., face images, medical data), 
just like the target organization. 
By experiments using 
two real 
datasets, we demonstrate that
our new technique enables such an adversary to perform the membership inference attack against transfer learning with high accuracy.

\noindent{\textbf{Contributions.}}
Our main contributions are as follows:

\begin{itemize}
\item We propose \TransMIA{} (Transfer learning-based Membership Inference Attacks), new 
attacks 
that 
use transfer learning to perform membership inference attacks on the source model. 
In particular,
we propose a \emph{transfer shadow training} technique, which transfers the parameters of the transferred model to a shadow model, to significantly improve the accuracy of membership inference when the amount of shadow training data is small. 
To our knowledge, this work is the first 
to use transfer learning to strengthen privacy attacks on the source model.
Zou~\etal{}~\cite{Zou_arXiv20} proposed a privacy attack against transfer learning (in a black-box access setting), whereas we propose a privacy attack using transfer learning (in a white-box access setting).
\item We evaluate our membership inference attacks by experiments using two real datasets,
and show that our membership inference attacks are realistic and effective threats to 
the source model 
especially when the adversary has 
a limited 
prior knowledge on the datasets.
Specifically, our attacks outperform 
a 
state-of-the-art 
method \cite{Shokri_SP17} 
that does not use our transfer shadow training technique.
\item We compare four types of our membership inference attacks to investigate the advantages and disadvantages of these attacks.
Specifically, 
we evaluate 
the four combinations of the learning-based/entropy-based approach and the fine-tuning/freezing approach,
all of which are 
trained with 
our transfer shadow training technique. 
Then we 
show 
that the learning-based approach has much higher accuracy than the entropy-based approach.
We also show that the attack with the fine-tuning approach may overfit shadow training datasets when the shadow training datasets are small.
Furthermore, we 
discuss the results of these four 
approaches 
on the basis of confidence values, and discuss possible countermeasures for our attacks.
\end{itemize}

\subsection{Related Work}
\label{sec:related}
\noindent{\textbf{Membership inference attacks.}}~~%
Shokri \etal{} \cite{Shokri_SP17} studied a membership inference attack which, given an individual data record and black-box access to a model, determines if the record was used to train the model. 
Here they proposed a shadow training technique, which trains shadow models that imitate the behavior of the target model. 
Inspired by \cite{Shokri_SP17}, a number of studies have been made on this attack \cite{Hayes_PoPETs19,Hilprecht_PoPETs19,Song_CCS19,Li:corr:20,Pyrgelis_NDSS18,Nasr_SP19,Jayaraman_arXiv20}. 
For example, Jayaraman \etal{} \cite{Jayaraman_arXiv20} assumed a realistic scenario where the candidate pool from which the adversary samples records is imbalanced, and 
proposed a membership inference attack based on the direction of change in loss of a query record when it is perturbed with a small amount of noise. 
They 
showed that their attack outperforms previous attacks in imbalanced prior settings. 
However, their attack still considers only the black-box access to the model.

Nasr \etal{} \cite{Nasr_SP19} evaluated white-box membership inference attacks against deep learning algorithms. They first showed that a straightforward extension of the black-box attack to the white-box setting, which uses the outputs of individual layers (rather than the outputs of the last layers), does not provide better accuracy than the black-box attack. 
Then they proposed a white-box attack using the gradient of the prediction loss with regard to all parameters, and showed that 
it 
outperforms the black-box attack. 
Their proposed attack cannot be used in our setting where the adversary does not have a white-box access to $\hT$, because the adversary cannot calculate the gradient by the back-propagation algorithm in this case. 

Recently, Song \textit{et al.} \cite{Song_arXiv20} proposed a membership inference attack based on modified prediction entropy (MPE). 
They showed that the MPE-based attack outperforms the DNN-based attack in their experiments. 
On the other hand, the MPE-based attack results in low accuracy in our experiments, because the distributions of confidence values are similar between training and test datasets and a single threshold does not separate the two distributions.

\smallskip
\noindent{\textbf{Defences against membership inference attacks.}}~~Defences against membership inference attacks have also been widely studied; 
e.g., regularization-based defenses \cite{Nasr:18:CCS,Salem:19:NDSS,Shokri_SP17}, DP (Differential Privacy)-based defenses \cite{Abadi_CCS16,Papernot_ICLR17,Shokri_CCS15,Yu_SP19}, 
adding noise to (or transforming) a confidence score vector \cite{Jia:19:CCS,Yang:corr:20:2005-03915}, 
and DAMIA \cite{Huang:corr:20:2005-08016}. 
The regularization-based defenses use regularization techniques to reduce overfitting; e.g., $l_2$-norm regularization \cite{Shokri_SP17}, dropout \cite{Salem:19:NDSS}, and adversarial regularization \cite{Nasr:18:CCS}. 
The DP-based defenses aim to prevent membership inference attacks from adversaries with any background knowledge. 
The regularization-based defenses and the DP-based defenses have no formal utility-loss guarantees, and 
can decrease the utility when we want to achieve high privacy. 
In the DP-based defenses, 
the privacy budget $\epsilon$ tends to be large to achieve high utility. 
For example, 
Rahman \etal{} \cite{Rahman_TDP18} reported that centralized deep learning in \cite{Abadi_CCS16} requires $\epsilon \geq 4$ to achieve high utility, and 
that it is vulnerable to membership inference attacks in this case.

Jia \etal{} \cite{Jia:19:CCS} proposed MemGuard, which adds noise to a confidence score vector based on adversarial example. 
Their defense bounds the loss of utility and provides better privacy-utility trade-off than the state-of-the-art regularization-based defenses and DP-based defenses. 
However, this defense cannot be applied to the whole parameters in the transferred model $\gS$. 
The same applies to the defense in \cite{Yang:corr:20:2005-03915}.

Finally, Huang \etal{} \cite{Huang:corr:20:2005-08016} proposed DAMIA based on DA (Domain Adaptation), 
which uses non-sensitive training data from a source domain to obfuscate sensitive training data from a target domain. 
Specifically, it works as follows: 
(i) remove labels of sensitive training data; 
(ii) train a model via DA. 
In step (ii), it uses as input sensitive training data without labels and non-sensitive training data with labels. 

Our work totally differs from \cite{Huang:corr:20:2005-08016} in that 
ours uses transfer learning to increase the performance of the membership inference attack. 
Another crucial difference lies in source and target tasks; 
since DA assumes that the source and target tasks are the same \cite{Pan_TKDE10}, DAMIA is limited to the setting where two datasets are collected for the same task (e.g., 
two-class classification of cats and dogs). 
In contrast, 
the source and target tasks can be different in our setting;
e.g., 
two datasets can be used for classification of different image labels. 

\smallskip
\noindent{\textbf{Transfer learning.}}~~Transfer learning has been widely studied over the past few decades 
(see \cite{Pan_TKDE10,Weiss:16:JBD,Tan_arXiv18} for surveys). 
Privacy-preserving transfer learning has also been studied in recent years; e.g., 
a method based on homomorphic encryption and secret sharing \cite{Gao_BigData19}, methods based on DP \cite{Wang_ECMLPKDD18,Guo_arXiv18,Yao_IJCAI19,Papernot_ICLR17}.
 
However, no studies have used transfer learning to strengthen privacy attacks on the source model, to our knowledge. 
We proposed a transfer shadow training technique, and showed that transfer learning can be used for such purposes when the amount of shadow training data is limited.

\section{Preliminaries}
\label{sec:preliminaries}

\subsection{Basic Notations and Machine Learning}
\label{sub:ML}

In this paper we deal with supervised learning models with classification tasks.
We define an \emph{input space} $\calx$ as a set of feature vectors, 
an \emph{output space} $\caly$ as a set of class labels. 
Let 
$P$ be the distribution of all data points over $\calx\times\caly$.
Then a learning algorithm seeks a function that maps a feature vector $x\in\calx$ to its class label $y\in\caly$.
We denote by $\Dists\caly$ the set of all distributions over $\caly$.
Then each distribution in $\Dists\caly$ can be represented as a real vector $(p_1, \ldots, p_{|\caly|})$ with $\sum_{i=1}^{|\caly|} p_i = 1$.

A \emph{classification model} is a function $f: \calx \rightarrow \Dists\caly$ that maps a feature vector $x$ to 
a \emph{prediction vector} $f(x)$ that
associates each class label $y\in\caly$ with a classification confidence value $f(x)_y \in [0, 1]$ such that $\sum_{y\in\caly} f(x)_y = 1$.
The final classification result is the class label with the largest confidence value, i.e., $\argmax_{\hat{y}} f(x)_{\hat{y}}$.

To evaluate how badly the classification model $f$ fits each data point $(x, y)$, 
we employ a real-valued \emph{loss function} $l$ (e.g., cross entropy).
Given a \emph{training dataset} $\Dtrain \subseteq \calx\times\caly$, the training aims to find a classification model $f$ that minimizes the expected loss 
$L_{\Dtrain}(f) = \expect_{(x, y) \sim \Dtrain}[l(f(x), y)]$.
To evaluate the model $f$, a \emph{test dataset} $\Dtest \subseteq \calx\times\caly$ is used to compute 
$L_{\Dtest}(f) = \expect_{(x, y) \sim \Dtest}[l(f(x), y)]$.

\subsection{Deep Transfer Learning}
\label{sub:TL}

\begin{figure}[t]
\centering
\includegraphics[width=0.88\linewidth]{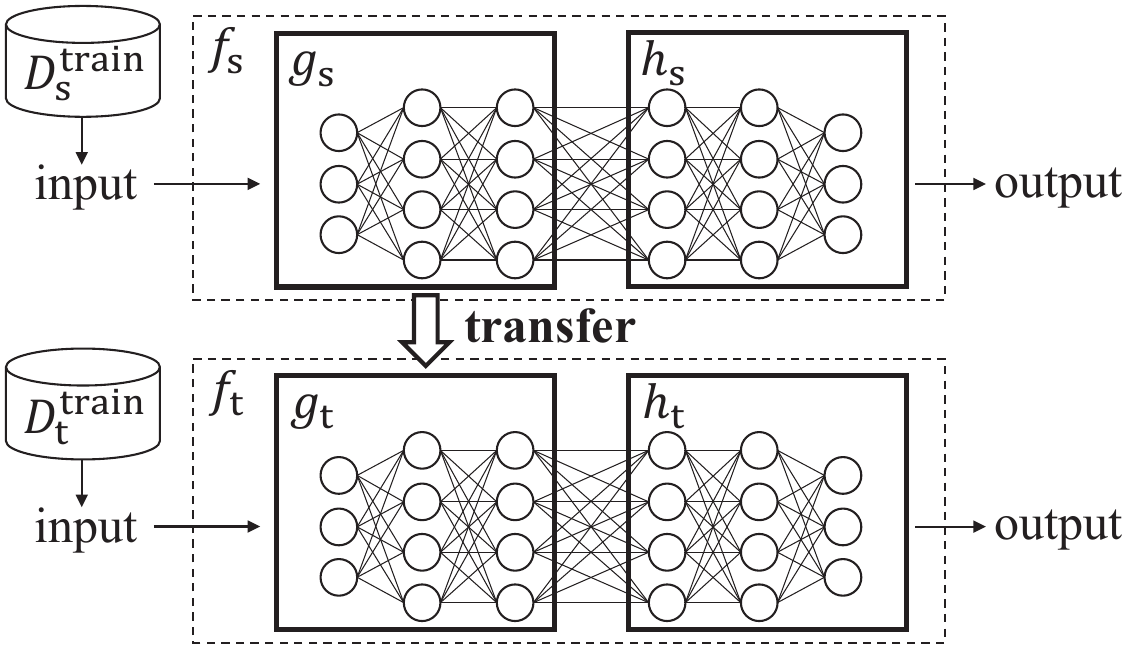}
\caption{Deep transfer learning. 
After a source model $\fS$ is trained using a dataset $\DStrain$,
the freezing approach initializes $\gT$ as $\gS$ and trains only $\hT$ using another dataset $\DTtrain$ ($\gT=\gS$). 
In contrast, the fine-tuning approach initializes $\gT$ as $\gS$ and updates the whole network ($\gT \neq \gS$).%
}
\label{fig:DTL}
\end{figure}

\emph{Transfer learning}~\cite{Caruana:94:NIPS,Bengio:11:JMLR,Bengio:12:JMLR,Pan_TKDE10,Weiss:16:JBD,Tan_arXiv18} is an approach to improving the performance of a machine learning task on a \emph{target domain} by transferring some knowledge from a \emph{source domain}.
\emph{Network-based deep transfer learning} \cite{Bengio:11:JMLR,Bengio:12:JMLR,Tan_arXiv18} is a branch of transfer learning that pre-trains a deep neural network $\fS$ from a large training dataset $\DStrain$ in the source domain and reuses a part of $\fS$ to construct another network $\fT$ 
with 
a significantly small dataset $\DTtrain$ in the target domain.
See \Fig{fig:DTL} for an  overview.

Formally, we define network-based deep transfer learning as follows.
Let $\calx$ be an \emph{input space},
$\calyS$ (resp.~$\calyT$) be a \emph{source (resp.~target) output space},
and $\PS$ (resp.~$\PT$) be the distribution of the data points over $\calx\times\calyS$ (resp.~$\calx\times\calyT$).
Then a \emph{source classification model} is a deep neural network $\fS: \calx\rightarrow\Dists\calyS$ that outputs a prediction vector $v$ providing each label $y$ with a classification confidence value $v_y\in [0,1]$,
and the \emph{source task} is defined as $\tS = (\calyS, \fS)$.
As illustrated in \Fig{fig:DTL}, a source model $\fS$ can be regarded as the cascade of the two networks $\gS$ and $\hS$ corresponding to the shallow and deep layers; i.e., $\fS = \hS \circ \gS$.
The shallow layers $\gS$ extract task-independent low-level features (e.g., edges, shapes), while the deep layers $\hS$ represent task-specific high-level features (e.g., eyes, mouths).
Analogously to the source domain, we also define a \emph{target model} $\fT = \hT \circ \gT$ and a \emph{target task} $\tT = (\calyT, \fT)$. 

In network-based deep transfer learning, a source model $\fS$ is first trained using a dataset $\DStrain \subseteq \calx\times\calyS$ and then its shallow layers $\gS$ are reused in the training of a target model $\fT$ 
with 
another dataset $\DTtrain \subseteq \calx\times\calyT$
(See \Fig{fig:DTL}).
We call $\DStrain$ a \emph{source training dataset}, and $\DTtrain$ a \emph{target training dataset}.
In later sections we also use a \emph{source test dataset} $\DStest \subseteq \calx\times\calyS$ and a \emph{target test dataset} $\DTtest \subseteq \calx\times\calyT$.

In this work we deal with two approaches: \emph{freezing} and \emph{fine-tuning}. 
In the freezing approach, we initialize and fix the shallow layers $\gT$ as $\gS$ and update only the deep layers in the training of the target model $\fT$.
Then the training aims to find a $\hT$ that minimizes the expected loss $L_{\PT}(\hT \circ \gS)$.
In contrast, in the fine-tuning approach, we initialize the shallow layers $\gT$ as $\gS$ and update the whole network in the training of $\fT$.
Then the training aims to find a $\hT \circ \gT$ that minimizes $L_{\PT}(\hT \circ \gT)$.
Empirically, the fine-tuning approach has better accuracy when target task labels are plentiful, whereas the freezing approach is better for scarce target task labels~\cite{Agrawal:14:ECCV,Girshick:14:CVPR}.

\subsection{Membership Inference Attack}
\label{sub:MIA}

We next review a privacy attack on machine learning models, called a \emph{membership inference attack} (MIA)~\cite{Shokri_SP17},
where an adversary is not given a training dataset $\Dtrain$ and attempts to reveal membership information on $\Dtrain$.
Specifically, the adversary in this attack aims to determine whether a specific data point $(x, y)$ is included in a training dataset $\Dtrain$ used to build a classification model $f$.

The membership inference attack exploits a different behavior of the model $f$ when a given data point $(x, y)$ has been used to train $f$.
To build an adversary $\Atk$ against $f$, we attempt to learn some statistical relevance between the distribution $f(x)$ of 
confidence values and the membership $(x,y)\in\Dtrain$.

In this attack, the adversary $\Atk$ 
is provided access to the model $f$
and some dataset disjoint from the training dataset $\Dtrain$.
An adversary is said to have \emph{black-box access} to $f$ if it can query data $x$ to $f$ and obtain their prediction vectors $f(x)$.
In contrast, \emph{white-box access} to $f$ allows the adversary to obtain the internal structure of $f$ itself (e.g., the weights of the connections between nodes when $f$ is 
a neural network).

Formally, an adversary in the black-box membership inference attack on a model $f$ is a function $\Atk: \calx\times\caly\times\Dists\caly \rightarrow \{\Lin, \Lout\}$ that given a data point $(x, y)$ and its prediction vector $f(x)$, outputs one of the labels $\Lin$ and $\Lout$,
where $\Lin$ (resp. $\Lout$) represents that $\Atk$ predicts $(x,y)\in\Dtrain$ (resp. $(x,y)\not\in\Dtrain$).
To evaluate the performance of the membership inference attack, we 
measure the \emph{accuracy}, \emph{precision}, and \emph{recall} of the attack by using
a test dataset $\Dtest \subseteq \calx\times\caly$ disjoint from the training dataset $\Dtrain$.

We deal with two approaches to constructing a membership inference adversary: the \emph{learning-based approach} \cite{Shokri_SP17} 
and the \emph{entropy-based approach} \cite{Song_arXiv20}.
The former constructs an adversary $\Atk$ as a classification model obtained by supervised learning using a dataset other than $\Dtrain$.
The latter approach calculates a variant of the Shannon entropy of the prediction vector (defined in Section~\ref{sub:modified:entropy}) and determine the membership by checking whether this entropy exceeds a certain threshold.

\subsection{Modified Prediction Entropy}
\label{sub:modified:entropy}

Finally, we review a recently proposed measure called \emph{modified prediction entropy}~\cite{Song_arXiv20},
which can be used to build a membership inference attack that may outperform the existing DNN-based attacks in some settings shown in \cite{Song_arXiv20}.
Intuitively, this measure is a variant of the Shannon entropy of a prediction vector $f(x)$ that also 
takes into account
the correct class label $y$.
Formally, 
the \emph{modified prediction entropy} $\ME(v, y)$ of a prediction vector $v\in\Dists\caly$ w.r.t. a label $y\in\caly$ is 
defined~by:
\begin{align*}
\ME(v, y) &= 
-(1 - v_y) \log(v_y)
-{\textstyle\sum_{y' \neq y}} v_{y'} \log(1 - v_{y'})
{.}
\end{align*}

\arxiv{
Then $\ME(f(x), y)$ monotonically decreases with the confidence value $f(x)_y$ of the correct label $y$, and monotonically increases with the confidence values $f(x)_{y'}$ of any incorrect labels $y'$.
When $f(x)_y = 1$ holds for the correct label $y$, the modified prediction entropy is $0$.
When $f(x)_{y'} = 1$ holds for an incorrect label $y'$, the modified prediction entropy is infinite.
}

\arxiv{
An entropy-based adversary of the membership inference attack is defined as a function that returns the label $\Lout$ if $\ME(f(x), y)$ is larger than some threshold 
$\tau_y \in \realsnng$
and returns $\Lin$ otherwise. 
The threshold $\tau_y$ is learned with the shadow training technique \cite{Song_arXiv20}. 
For example, $\tau_y$ is determined so that it maximizes the prediction accuracy for the shadow training dataset $\DSHtrain$ and shadow test dataset $\DSHtest$.
}

\section{Attack Scenarios}
\label{sec:illustrative}

We consider a small startup company (or small hospital) $\cstart$ that has only small datasets of images (e.g., face images, X-ray images) and wants to achieve higher performance of image classification.
Then a big company (or large hospital) $\cbig$, 
that 
possesses large datasets of images, provides a part of their ML model to the startup $\cstart$.
Specifically, 
$\cbig$ pre-trains a deep neural network $\fS$ from a large training dataset $\DStrain$ in the source domain, and the startup $\cstart$ reuses the shallow layers $\gS$ of 
$\fS$ to construct another network $\fT$ using a significantly small dataset $\DTtrain$ in the target domain. 
The big company $\cbig$ also offers a machine learning as a service (MLaaS) to general consumers 
via a black-box API. 
This service enables the general consumers 
to query an arbitrary input to obtain $\fS$'s output.
Note that the big company $\cbig$ does not allow white-box access to the whole source model $\fS$ to the general consumers, 
because otherwise a lot of membership information can be leaked by the parameters of $\fS$~\cite{Nasr_SP19}.

\begin{figure}[t]
\centering
\includegraphics[width=0.88\linewidth]{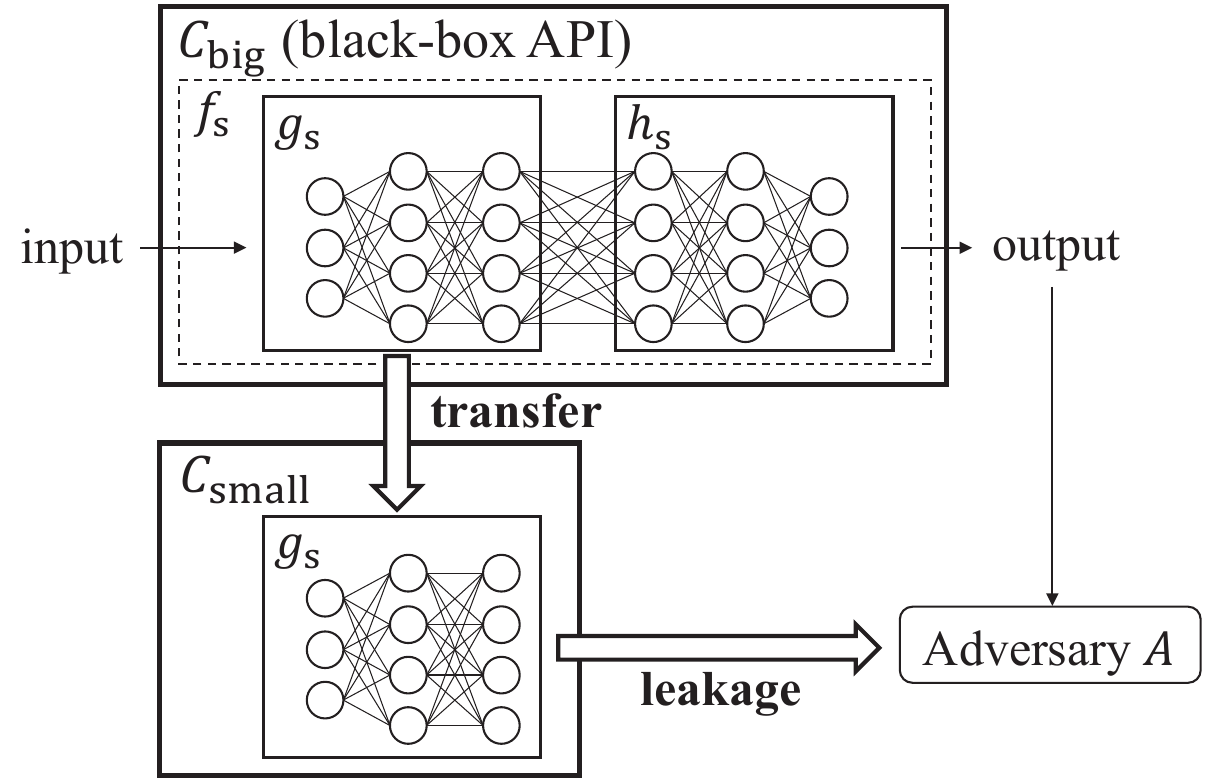}
\caption{Our attack scenario. After the big company $\cbig$ transfers the shallow layers $\gS$ to the small startup company $\cstart$, the adversary $\Atk$ obtains $\gS$ from $\cstart$ (e.g., via illegal access). The adversary $\Atk$ also obtains the output of 
$\fS$ via the black-box API provided by $\cbig$. The adversary $\Atk$ performs membership inference attack on $\fS$ using these information.}
\label{fig:first:scenario}
\end{figure}

Our attack scenario is that the startup company $\cstart$, which obtains the transferred model $\gS$ by signing a contract with the big company $\cbig$, leaks the shallow layers $\gS$ e.g., by illegal access, malware infection, or internal fraud. 
Note that even if the startup company $\cstart$ trains $\fT$ with the fine-tuning approach, $\gS$ can be leaked by $\cstart$. 
Then we consider an adversary $\Atk$ who has 
black-box access to the deep neural network $\fS$ through 
an 
API and 
has all the parameters of the shallow layers $\gS$
for the purpose of transfer learning (\Fig{fig:first:scenario}).
The adversary $\Atk$ 
performs a black-box membership inference attack on $\fS$ to learn whether a specific sensitive image $x$ is included in the training dataset $\DStrain$.
Although $\gS$ alone extracts only task-independent low-level features and may not directly be used to construct membership inference attacks, we show that $\gS$ can be used to strengthen the shadow training to mount a stronger adversary against $\fS$.
This implies that transfer learning can 
increase the vulnerability to the privacy breach from the source model $\fS$.

\section{Membership Inference Attacks Based on the Transfer Shadow Training Technique}
\label{sec:MIA-TST}
In this section we propose 
new membership inference attacks 
\TransMIA{}
where attackers perform transfer learning
in the attack scenarios in Section~\ref{sec:illustrative}.
In particular, we propose a technique called \emph{transfer shadow training} to significantly improve the performance of the
attacks when only a small amount of shadow training data is available to the adversary.

\subsection{Threat Model}
\label{sub:attack:threat}

We investigate a membership inference attack against a deep neural network $\fS$ that consists of shallow layers $\gS$ and deep layers $\hS$; i.e., $\fS = \hS \circ \gS$. (See \Fig{fig:DTL}.)
This attack aims to determine whether a specific data point $(x, y)$ is included in the source training dataset $\DStrain$.
In Section~\ref{sub:MI-transfer}, we construct an adversary $\Atk$ 
who 
has (1) black-box access to the network $\fS$ through an API, (2) white-box access to the shallow layers $\gS$ for the purpose of transfer learning, (3) no access to the deep layers $\hS$, and (4) may not necessarily know $\hS$'s underlying architecture or hyper parameters. 
In our experiments in Section~\ref{sec:evaluation}, we construct shadow models that have the same architecture as $\hS$. 
However, as shown in previous work such as~\cite{Shokri_SP17}, the knowledge of the architecture is not necessary to construct shadow models.

We assume that the adversary $\Atk$ possesses a shadow training dataset $\DSHtrain \subseteq \calx\times\calyS$ to train (possibly multiple) shadow models. 
The elements of $\DSHtrain$ are
drawn from the same distribution $\PS$ as the source training dataset $\DStrain$. 
Since the goal of the attack is to reveal membership information on $\DStrain$, the adversary $\Atk$ has no prior knowledge of $\DStrain$ itself, hence of whether a data point $(x, y)\in\DSHtrain$ is included in $\DStrain$ or not.
Typically, $\DSHtrain$ is disjoint from $\DStrain$.
Note that such a threat model is also assumed in most of the previous work on membership inference attacks (e.g.,~\cite{Shokri_SP17}).

We also assume that the adversary $\Atk$ possesses a shadow test dataset $\DSHtest\subseteq\calx\times\calyS$
disjoint from $\DSHtrain$ and $\DStest$. The adversary $\Atk$ uses the shadow training dataset $\DSHtrain$ and the shadow test dataset $\DSHtest$ to train an attack model.

Finally, we assume that 
both the shadow datasets $\DSHtrain$ and $\DSHtest$ consist 
of a small amount of data.
This assumption is reasonable, because it is expensive or difficult for the adversary to acquire a large amount of personal data, 
(e.g., face images, medical data), 
just like the small startup company $\cstart$. 
When $\DSHtrain$ is 
small, it is challenging for the adversary to perform membership inference attacks with high accuracy. 
To address this issue, we propose a technique called \emph{transfer shadow training}, which we describe below.

\subsection{Transfer Shadow Training}
\label{sub:MI-transfer_overview}
We now propose a \emph{transfer shadow training} technique. 
\begin{figure}[t]
\centering
\includegraphics[width=0.88\linewidth]{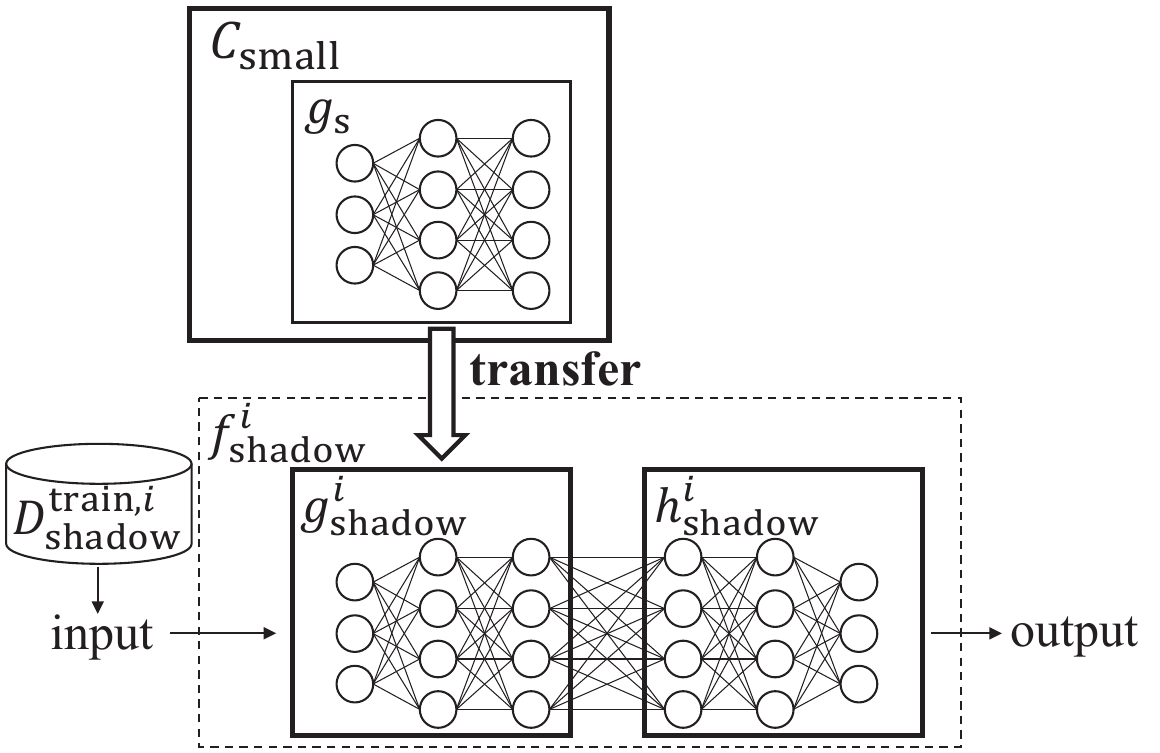}
\caption{Overview of the transfer shadow training.}
\label{fig:transfer_shadow_training}
\end{figure}
As shown in \Fig{fig:transfer_shadow_training},
the adversary $\Atk$ obtains the shallow layers $\gS$ from the small startup company $\cstart$ (e.g., via illegal access), as described in Section~\ref{sec:illustrative}. 
Then the adversary $\Atk$ transfers this information to shadow models to obtain stronger attack performance with a small amount of shadow training data.

Specifically, we train multiple \emph{shadow models} 
to simulate the behavior of the original classification model $\fS$ under attacks. 
We denote by $\cSH \in \nats$ the number of the shadow models.
For each $i = 1, 2, \ldots , \cSH$, let $\fSH^i$ be the $i$-th shadow model. 
We regard $\fSH^i$ as a cascade of shallow layers $\gSH^i$ and deep layers $\hSH^i$; i.e., $\fSH^i = \hSH^i \circ \gSH^i$. 
We use a dataset $\DSHItrain \subseteq \DSHtrain$ to train the $i$-th shadow model $\fSH^i$, and refer to it as the \emph{shadow training dataset for $\fSH^i$}.
As with \cite{Shokri_SP17}, the training datasets for different shadow models  may intersect.

To construct each shadow model $\fSH^i$, our transfer shadow training technique initializes the shallow layers $\gSH^i$ as $\gS$. 
Then it trains the shadow model $\fSH^i$ by using the dataset $\DSHItrain$. 
In other words, we perform the network-based deep transfer learning to construct the shadow model.
As we will show in our experiments, this significantly improves the accuracy of membership inference even when the amount of shadow training data $\DSHtrain$ is small.
In the transfer shadow training we may take the freezing approach ($\gSH^i = \gS$) or the fine-tuning approach ($\gSH^i \allowbreak \neq \gS$).

\subsection{Attack Algorithms}
\label{sub:MI-transfer}

Next we describe algorithms for \TransMIA{} (Transfer learning-based Membership Inference Attacks), membership inference attacks using our transfer shadow training technique.
Here we deal with two types of adversaries: \emph{learning-based} and \emph{entropy-based}.
This is because in a recent work \cite{Song_arXiv20}, entropy-based membership inference attacks are proposed and shown to outperform the DNN-based membership inference attack in their experiments.
We denote our membership inference attacks with the learning-based approach and the entropy-based approach by \TransMIAL{} and \TransMIAE{}, respectively. 
To construct the attack algorithms, we introduce a shadow test dataset $\DSHItest \subseteq \DSHtest$ for each shadow model $\fSH^i$.
As with the shadow training datasets, the test datasets for different shadow models may also intersect.

First, we construct a learning-based adversary using our transfer shadow training technique
in the following three steps.
\\\\ 
\underline{\textbf{\TransMIAL{} (learning-based adversary)}}
\begin{enumerate}
\item[(L1)]\, We first train multiple \emph{shadow models} $\fSH^i$ for $i = 1, 2, \ldots , \cSH$. 
Specifically, we initialize the shallow layers $\gSH^i$ as $\gS$, and then train $\fSH^i$ by using the dataset $\DSHItrain \subseteq \DSHtrain$.
\item[(L2)]\, We next obtain a dataset $\DAtrain$ labeled with membership information ($\Lin$ or $\Lout$) by applying each shadow model $\fSH^i$ to the shadow training dataset $\DSHItrain$ and the shadow test dataset $\DSHItest \subseteq \DSHtest$. 
Specifically, we define an \emph{attack training dataset} $\DAtrain$ as the set of the records $(x, y, \allowbreak \fSH^i(x), \Lin)$ for each $(x, y)\in\DSHItrain$ and $(x', y', \fSH^i(x'), \allowbreak \Lout)$ for each $(x', y')\in\DSHItest$ 
where $i = 1, 2, \ldots , \cSH$.

\item[(L3)]\, We finally construct a membership inference adversary model $\Atk: \calx\times\calyS\times\Dists\calyS \rightarrow \{\Lin, \Lout\}$ as a classification model 
(e.g., DNN, SVM) 
using the attack training dataset $\DAtrain$ constructed in (L2).
\end{enumerate}
Given an input $(x,y)\in\calx\times\calyS$, the learning-based adversary \TransMIAL{} obtains $\fS(x)$ by black-box access to the source model $\fS$ and outputs the label $\Atk(x, y, \fS(x))$.

Next, we construct an entropy-based adversary with transfer shadow training in the following three steps.\\\\
\underline{\textbf{\TransMIAE{} (entropy-based adversary)}}
\begin{enumerate}
\item[(E1)]\, We first train multiple \emph{shadow models} $\fSH^i$ 
for $i = 1, 2, \ldots , \cSH$ in the same way as in (L1).

\item[(E2)]\, We next employ each shadow model $\fSH^i$ and calculate the modified prediction entropy $\ME(\fSH^i(x), y)$ for each 
$(x, y)\in\DSHItrain\cup\DSHItest$.
Then for each class label $y\in\calyS$, we select a threshold $\tau_y$ that maximizes:
\begin{align*}
& {\textstyle\sum_{i=1}^{\cSH}} |\{ (x, y)\in\DSHItrain : \ME(\fSH^i(x), y) \le \tau_y \}| 
\\[0.0ex]& \hspace{9mm} +
|\{ (x, y)\in\DSHItest : \ME(\fSH^i(x), y) > \tau_y \}|.
\end{align*}
Then $\tau_y$ maximizes the accuracy of the attack.

\item[(E3)]\, We finally define a membership inference adversary $\Atk: \calx\times\calyS\times\Dists\calyS \rightarrow \{\Lin, \Lout\}$ by
$\Atk(x, y, \fS(x)) = \Lin$ if $\ME(\fS(x), y) \le \tau_y$, and
$\Atk(x, y, \fS(x)) = \Lout$ otherwise.
\end{enumerate}
Given an input $(x,y)\in\calx\times\calyS$, the entropy-based adversary \TransMIAE{} obtains $\fS(x)$ by black-box access to the source model $\fS$ and outputs the label $\Atk(x, y, \fS(x))$.

In each of \TransMIAL{} and \TransMIAE{}, the transfer shadow training may take the freezing or fine-tuning approach. 
Therefore, there are four types of our membership inference attacks,
which we compare in our experiments.
As a membership inference adversary model $\Atk$ in \TransMIAL{}, we use a deep neural network (DNN) and a support vector machine (SVM).

\section{Evaluation}
\label{sec:evaluation}

\subsection{Datasets and Source Models}
\label{sub:datasets}
We conducted experiments to evaluate our attacks
by using two real datasets that are typically used as source datasets for experimental evaluation: 
\emph{ImageNet} 
and \emph{VGGFace2}~\cite{Cao2018}.
While several pre-trained classification models based on these datasets are provided, we construct the source models from scratch. This is because the exact source training datasets used for constructing the models are required for the evaluation.
In this section we explain these datasets and the source classification models on which we perform attacks.

\smallskip
\noindent{\textbf{ImageNet.}}~~%
\emph{ImageNet} is a large-scale image dataset designed for visual object recognition research.
We especially use the \emph{ImageNet Large Scale Visual Recognition Challenge dataset 2012} (\emph{ILSVRC2012})~\cite{ILSVRC15}, 
which is a trimmed ImageNet dataset consisting of 1,000 categories of images.
We randomly select 100 categories from all categories with more than 800 images.
We split the dataset into four disjoint datasets: a training dataset $\DStrain$, a test dataset $\DStest$, a shadow training dataset $\DSHtrain$, and a shadow test dataset $\DSHtest$.
Both $\DStrain$ and $\DStest$ consist of 50 images for each category, and 5,000 images in total.
The shadow datasets $\DSHtrain$ and $\DSHtest$ have 62,013 and 62,010 images, respectively.
We construct 
a classification model $\fS$ with ResNet18~\cite{He_arXiv15}.
We iterate the training of $\fS$ over 120 epochs, and save the updated model every epoch.
We finally select a model with the best accuracy from the saved models as $\fS$.
The accuracy of the model $\fS$ for $\DStrain$ and $\DStest$ are 0.298 and 0.242, respectively.

\smallskip
\noindent{\textbf{VGGFace2.}}~~%
\emph{VGGFace2} is a large-scale face recognition dataset
with more than 9,000 identities.
We randomly select 100 identities from the identities with more than 500 images.
As with the ImageNet dataset, we split the dataset into four disjoint datasets $\DStrain$, $\DStest$, $\DSHtrain$, and $\DSHtest$.
Both $\DStrain$ and $\DStest$ consist of 50 images for each identity, and 5,000 images in total.
The shadow datasets $\DSHtrain$ and $\DSHtest$ have  22,623 and 22,680 images, respectively.
We assume a task to predict the identity for a given face image.
We construct a model $\fS$ with ResNet18~\cite{He_arXiv15}.
We iterate the training of $\fS$ over 120 epochs, and select the best model in the same way as the ImageNet dataset.
The accuracy of the model $\fS$ for $\DStrain$ and $\DStest$ are 0.855 and 0.752, respectively.

\subsection{Experimental Setup}
\label{sub:setup}
In this section we describe the shadow models and attack models.
We assume a scenario where all the layers of the target model $\fS$ except the last fully connected layer are published as $\gS$ for transfer learning.
The adversary leverages $\gS$ for transfer shadow training.
We apply the same settings to both target models generated with the ImageNet and VGGFace2 datasets.

We construct shadow models with three learning approaches: \emph{baseline}, \emph{freezing}, and \emph{fine-tuning}.
The baseline approach,
denoted by \BaselineMIA{},
is the original membership inference attack proposed by Shokri~\etal{}~\cite{Shokri_SP17},
where we do not utilize any layers of the transferred model $\gS$ to construct shadow models. 
The freezing and fine-tuning approaches are instances of our proposed methods described in Section~\ref{sub:MI-transfer_overview}.

In our experiments, we construct 100 shadow models for each learning approach (i.e., $\cSH = 100$). 
For each shadow model $\fSH^i$, we prepare a shadow training dataset $\DSHItrain$ and a shadow test dataset $\DSHItest$ by randomly 
selecting images 
of $\DSHtrain$ and $\DSHtest$, respectively.
Each shadow model $\fSH^i$ is trained with 
the corresponding shadow training dataset $\DSHItrain$. 
In our experiments, the size of $\DSHItrain$ is
100, 200, 300, 400, 500, 600, 800, or 1,000.
We also construct a shadow test dataset $\DSHItest$ for $\fSH^i$ with the same size as $\DSHItrain$.
As a result, the attack training dataset $\DAtrain$ has the same number of records with the $\Lin$ and $\Lout$ labels. 

We perform 
our transfer learning-based membership inference attack with the \emph{learning-based approach} and the \emph{entropy-based approach}.
For learning-based approach, we construct two types of attack models: a deep neural network (DNN) and a support vector machine (SVM),
denoted by \TransMIADNN{} and \TransMIASVM{} respectively.
In our experiments, we evaluate \TransMIADNN{}, \TransMIASVM{}, and \TransMIAE{}.

The neural network has three fully-connected hidden layers.
The numbers of neurons for the hidden layers are 50, 30, and 5.
For each class label $y$, we obtain an attack training dataset $\DAtrain$ and construct the attack model  (i.e.,  100 attack models in total).
We train each attack model with 50 epochs.

We evaluate attack performance with \emph{accuracy}, \emph{precision}, and \emph{recall}.
For each dataset size, we construct 10 attack training datasets $\DAtrain$ while varying images selected from $\DSHtrain$ and $\DSHtest$.
We then perform each attack 10 times with different attack training datasets.
We average the results of 10 attacks for each performance metric.
We use $\DStrain$ and $\DStest$ as test datasets for the attack models.

\subsection{Experimental Results}
\label{sub:attacks:MIA}

We first 
present experimental results on 
our transfer learning-based 
attack \TransMIADNN{} with the DNN and compare it with the baseline approach \BaselineMIA{}.

\begin{figure}[t]
 \centering
 \begin{minipage}{0.475\hsize}
  \centering
  \includegraphics[width=1.0\hsize]{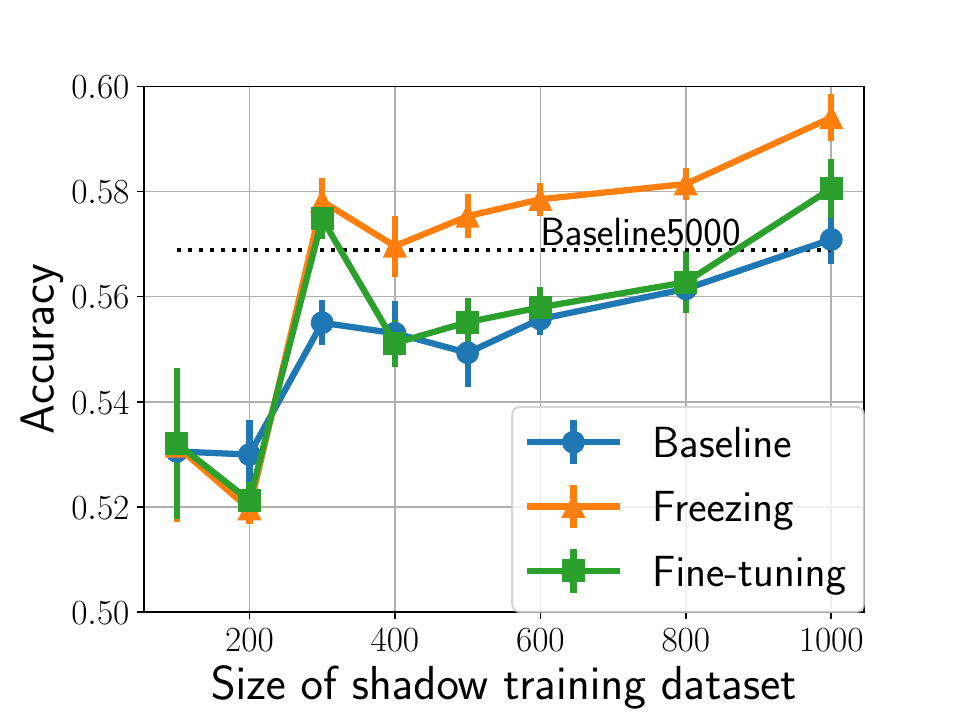}
  \subcaption*{(a) Accuracy in ImageNet.}
  \label{fig:imagenet_accuracy}
 \end{minipage}
 \begin{minipage}{0.475\hsize}
  \centering
  \includegraphics[width=1.0\hsize]{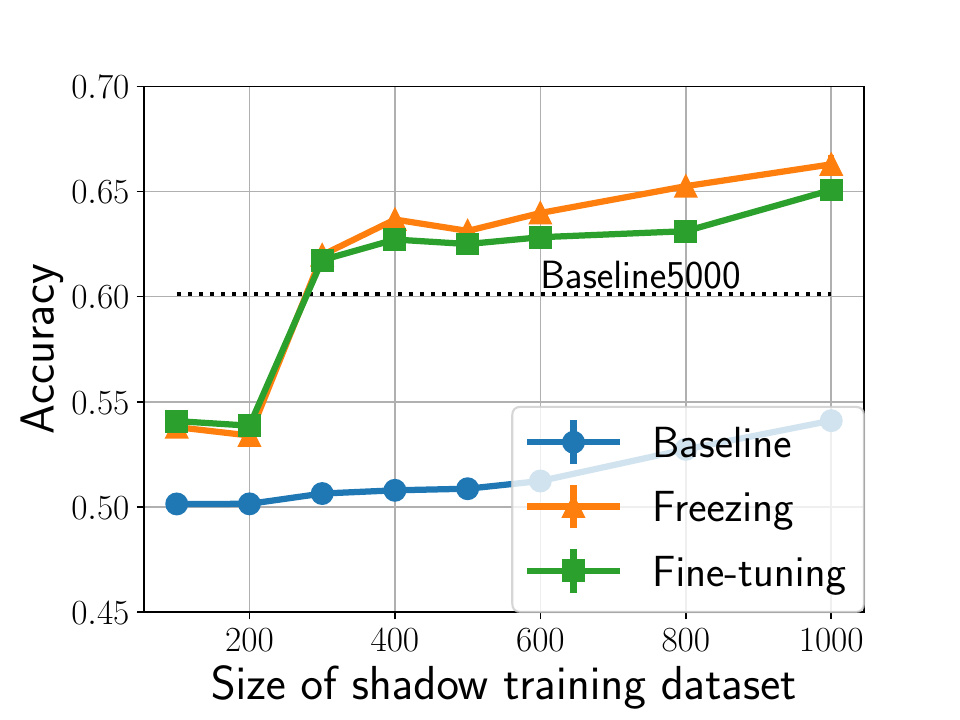}
  \subcaption*{(d) Accuracy in VGGFace2.}
  \label{fig:vggface2_accuracy}
 \end{minipage}
 \begin{minipage}{0.475\hsize}
  \centering
  \includegraphics[width=1.0\hsize]{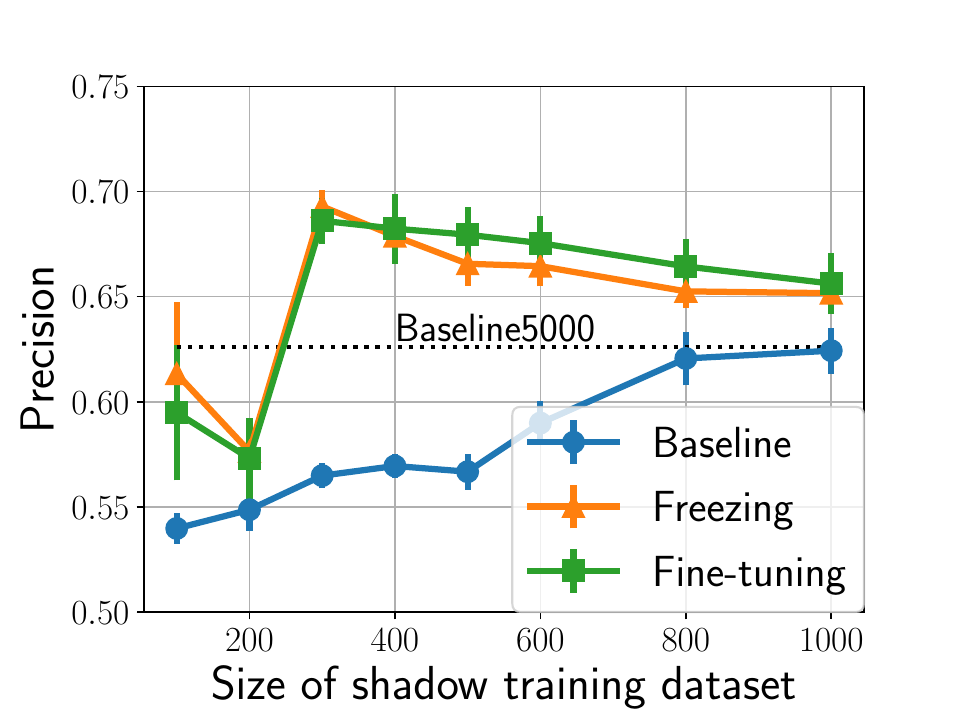}
  \subcaption*{(b) Precision in ImageNet.}
  \label{fig:imagenet_precision}
 \end{minipage}
 \begin{minipage}{0.475\hsize}
  \centering
  \includegraphics[width=1.0\hsize]{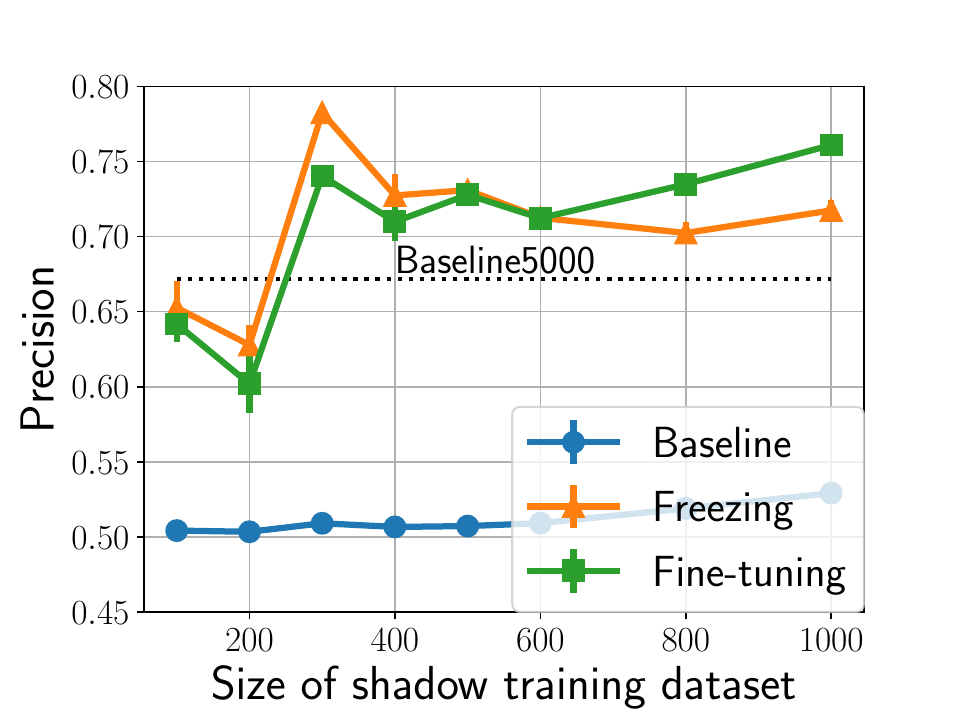}
  \subcaption*{(e) Precision in VGGFace2.}
  \label{fig:vggface2_precision}
 \end{minipage}
 \begin{minipage}{0.475\hsize}
  \centering
  \includegraphics[width=1.0\hsize]{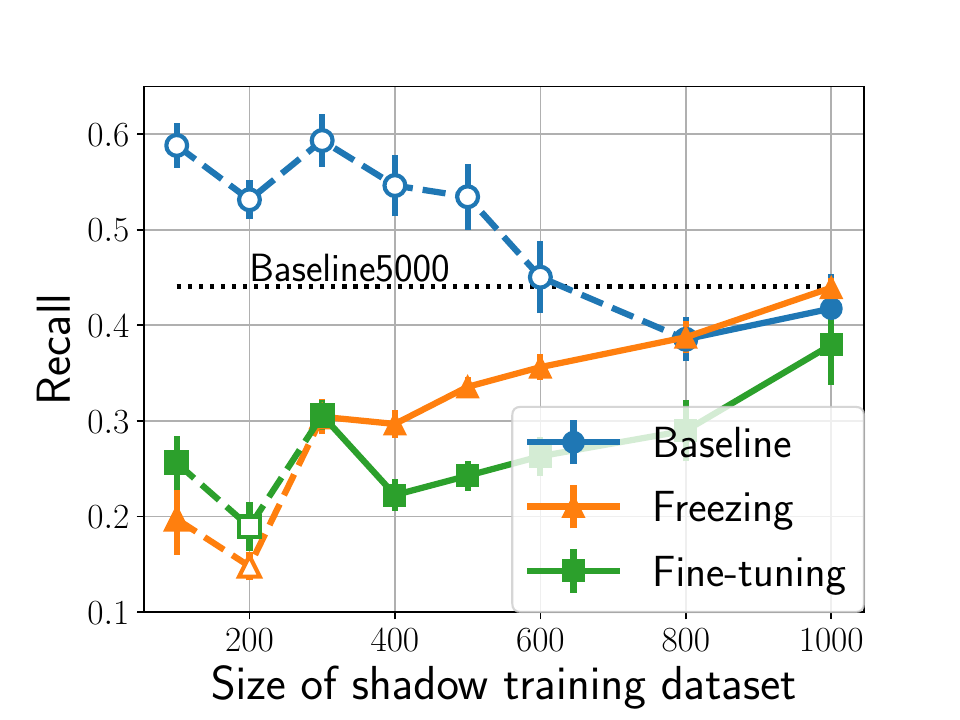}
  \subcaption*{(c) Recall in ImageNet.}
  \label{fig:imagenet_recall}
 \end{minipage}
  \begin{minipage}{0.475\hsize}
  \centering
  \includegraphics[width=1.0\hsize]{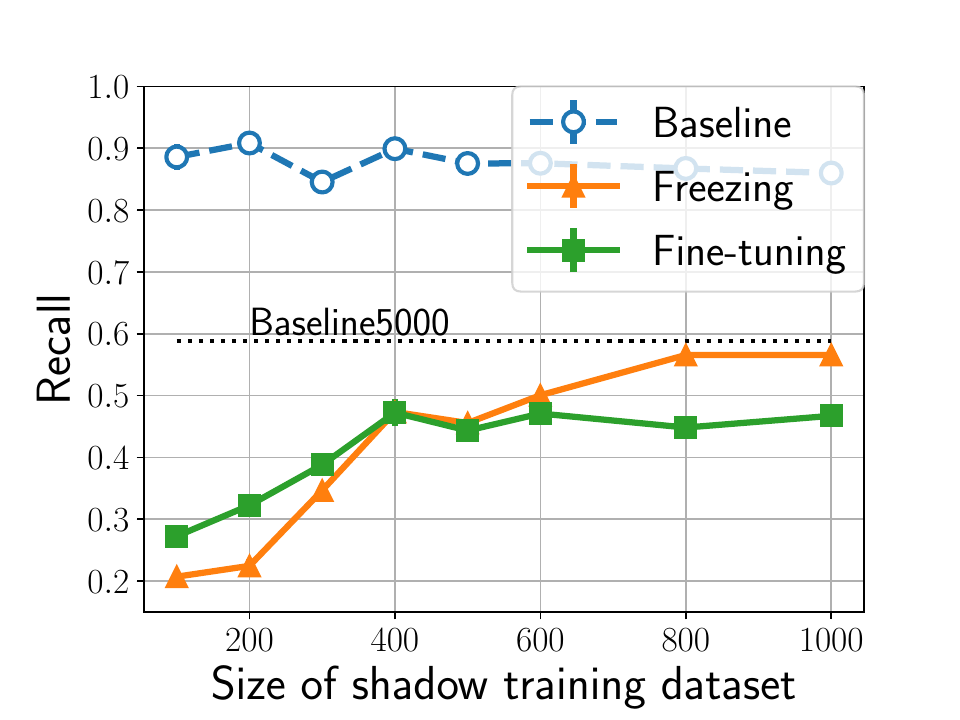}
  \subcaption*{(g) Recall in VGGFace2.}
  \label{fig:vggface2_recall}
 \end{minipage}
 \caption{Accuracy, precision, and recall of 
 the baseline attack \BaselineMIA{} and our attacks \TransMIADNN{} with the learning-based approach (DNN). 
 In \TransMIADNN{}, we evaluate both 
 freezing and fine-tuning approaches. 
 Figures (a)--(c) (resp. (d)--(f)) are the results in the ImageNet (resp. VGGFace2) dataset. Each line indicates the average of results for 100 classes. Error bars indicate the standard deviation of results on 10 attacks with different shadow training datasets. The dotted line annotated as ``Baseline5000'' indicates the result of 
 \BaselineMIA{} 
 when the size of the shadow training dataset is 5,000. The dashed lines and outline markers in Figures (c) and (f) indicate that the corresponding precision value is less than 60\%. }
 \label{fig:acc_prc_rec}
\end{figure}
\Fig{fig:acc_prc_rec} shows accuracy, precision, and recall. 
Here we evaluated three approaches: \BaselineMIA{}, \TransMIADNN{} with the freezing approach, \TransMIADNN{} with the fine-tuning approach. 
``Baseline5000'' 
(dotted line) 
indicates the result of 
\BaselineMIA{} 
when each shadow training dataset $\DSHItrain$ consists of 5,000 images, i.e., when the adversary utilizes shadow training datasets with the same size as 
$\DStrain$.

\Fig{fig:acc_prc_rec} shows that 
our 
attack \TransMIADNN{} 
with the freezing/fine-tuning approach 
achieves 
higher accuracy and precision than the attack with the baseline approach \BaselineMIA{}.
In addition, the accuracy and precision of our 
attack with the freezing approach 
are over the values of baseline5000 when the size of each shadow training dataset $\DSHItrain$ is greater than or equal to 300.
These results show that transfer shadow training is effective for the attack with a small shadow training datasets $\DSHItrain$.
In our experiments with the ImageNet dataset, the accuracy of the fine-tuning approach is worse than that of the freezing approach.
The reason is that the fine-tuning approach empirically tends to overfit the target training dataset when the dataset is small~\cite{Agrawal:14:ECCV,Girshick:14:CVPR}.

It should be noted that although the recall of the baseline approach \BaselineMIA{} is high at the small sizes, the corresponding precision 
is 
very low: 
less than 
0.6 
(and therefore, we show the recall in this case using the dashed line). 
When the precision is close to 0.5, the adversary fails to infer membership information, irrespective of the recall. 
For example, consider an adversary who randomly outputs ``$\Lin$'' (member) with probability $p \in [0,1]$ and ``$\Lout$'' (non-member) with probability $1-p$. 
The accuracy, precision, and recall of this random guess are 0.5, 0.5 and $p$, respectively. 
Thus, this random guess achieves recall $=1$ when $p=1$. 
However, it completely fails to infer membership information (in fact, precision is 0.5).
Similarly, \BaselineMIA{} outputs ``$\Lin$'' many times and this explains high recall of \BaselineMIA{}. 
However, \BaselineMIA{} behaves like the random guess with high $p$, and fails to infer membership information (precision is close to 0.5).

\Fig{fig:acc_prc_rec} shows that the recall of our membership inference attack \TransMIADNN{} is low at the small sizes. 
This is because shadow training datasets $\DSHItrain$ include few data points that are similar to source training data. 
However, the recall 
of \TransMIADNN{} 
increases with an increase in the size of the shadow training dataset.
When the size is 1,000, the recall of the freezing approach is close to the value of baseline5000.
Since the size of 1000 is lower than the full size (i.e., 5,000), 
our attack is effective in terms of recall as well.
Note that our attack also achieves high accuracy and precision; i.e., 
when the size is 1,000, 
\TransMIADNN{} with the freezing approach achieves higher accuracy and precision than baseline5000, and almost the same recall as baseline5000 in both the ImageNet and VGGFace2 datasets.

\begin{figure*}[t]
\begin{center}
 \begin{minipage}{0.245\hsize}
  \centering
  \includegraphics[width=1.0\hsize]{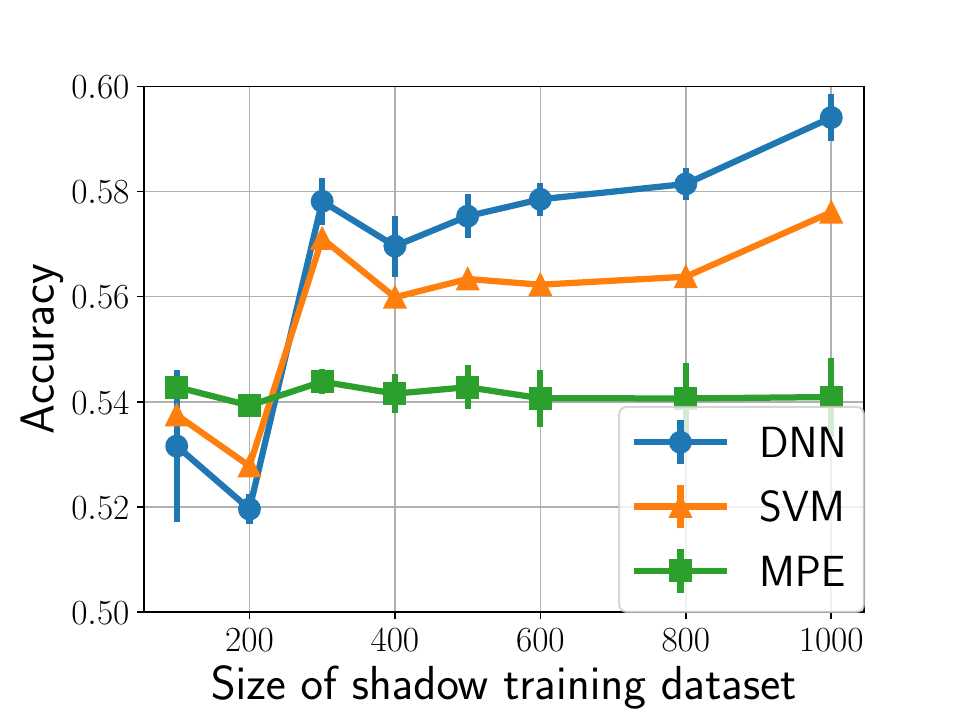}%
  \subcaption{Freezing for ImageNet.}
  \label{fig:imagenet_freezing}
 \end{minipage}
 \begin{minipage}{0.245\hsize}
  \centering
  \includegraphics[width=1.0\hsize]{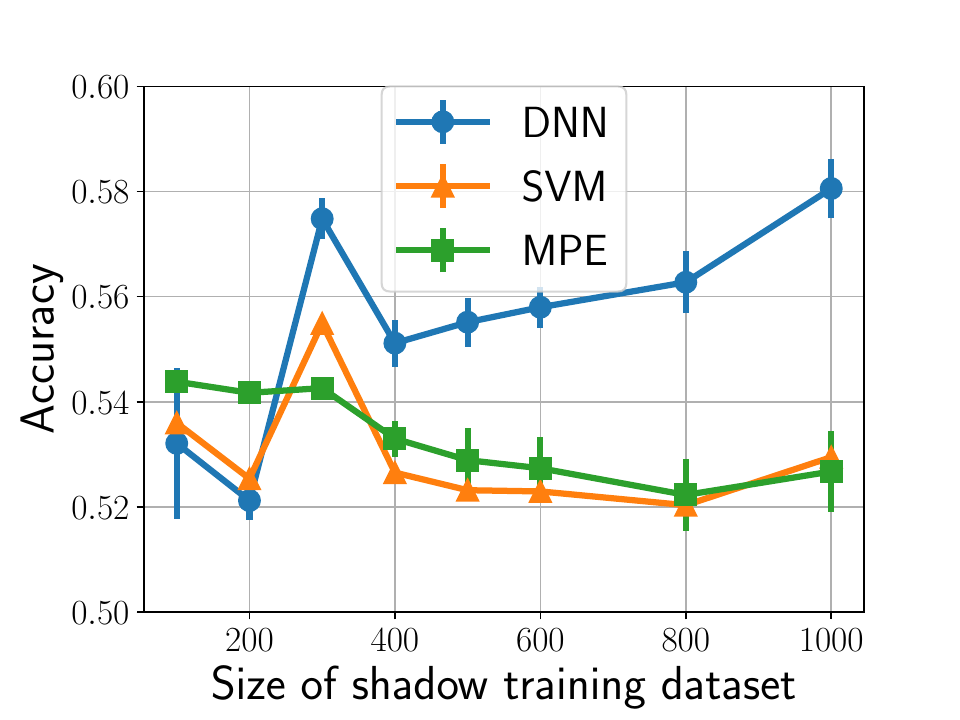}%
  \subcaption{Fine-tuning for ImageNet.}
  \label{fig:imagenet_finetuning}
 \end{minipage}
  \begin{minipage}{0.245\hsize}
  \centering
  \includegraphics[width=1.0\hsize]{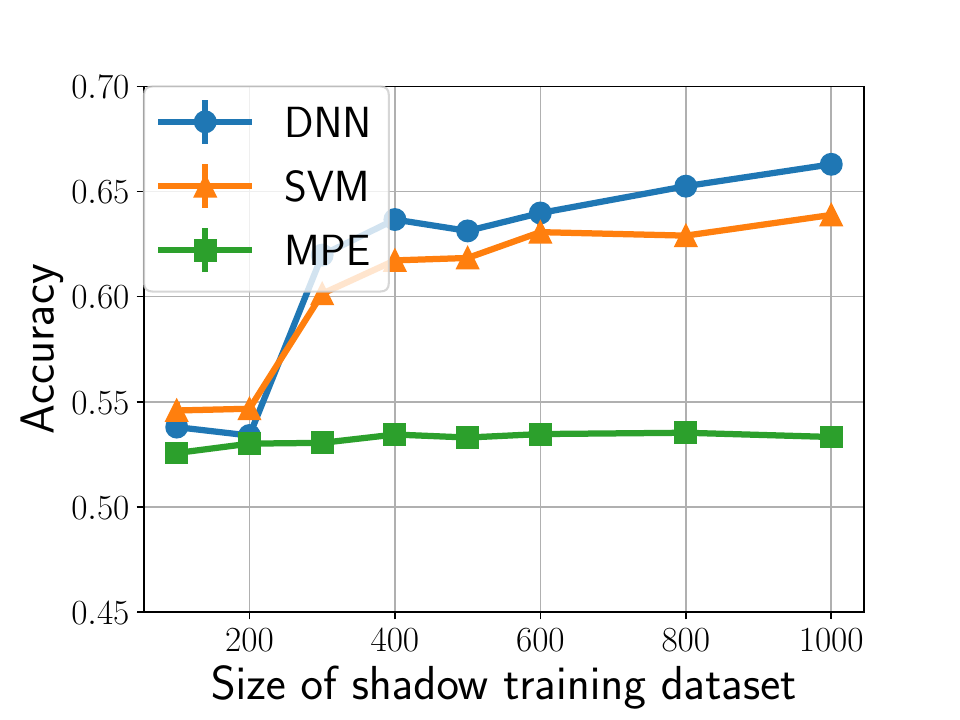}%
  \subcaption{Freezing for VGGFace2.}
  \label{fig:vggface2_freezing}
 \end{minipage}
 \begin{minipage}{0.245\hsize}
  \centering
  \includegraphics[width=1.0\hsize]{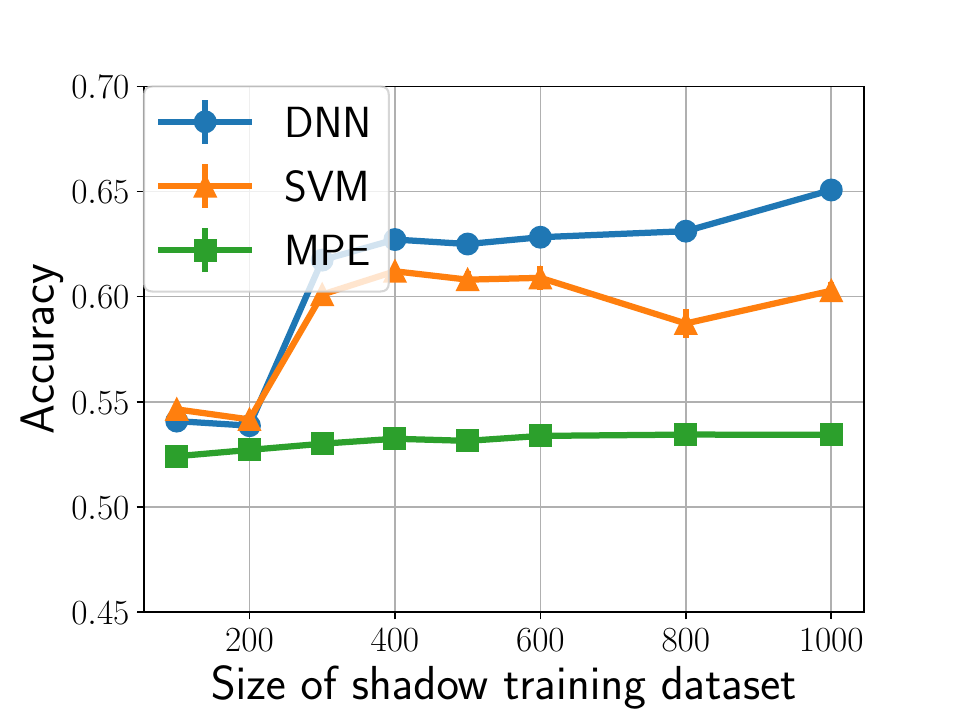}%
  \subcaption{Fine-tuning for VGGFace2.}
  \label{fig:vggface2_finetuning}
 \end{minipage}
 \caption{Accuracy of our attacks \TransMIADNN{}, \TransMIASVM{}, and \TransMIAE{} 
 with three different attack models: DNN, SVM, and MPE (modified prediction entropy). The left figures (a) and (c) shows our attacks with the freezing approach, whereas the right figures (b) and (d) shows our attacks with the fine-tuning approach. Figures (a) and (b) (resp. (c) and (d)) show the results in the ImageNet (resp. VGGFace2) dataset. As with \Fig{fig:acc_prc_rec}, each line indicates averages for 100 classes, and error bars indicate standard deviations for 10 attacks.}
 \label{fig:frz_fit}
\end{center}
\end{figure*}

We next present experimental results on our attacks with the DNN, SVM, and the entropy-based approach, i.e., \TransMIADNN{}, \TransMIASVM{}, and \TransMIAE{}. 
\Fig{fig:frz_fit} shows the accuracy of our attacks with the three different attack models. 
\Fig{fig:frz_fit} shows that the learning-based approaches \TransMIADNN{} and \TransMIASVM{}
have much higher accuracy than the entropy-based approach \TransMIAE{}.
However, 
\TransMIASVM{} with the fine-tuning approach 
results in 
low accuracy for the ImageNet dataset.
As 
explained 
above, each shadow 
model 
$\fSH^i$ constructed with the fine-tuning approach on the ImageNet dataset overfits shadow training 
dataset 
$\DSHItrain$.
In this case, the decision boundary between $\Lin$ and $\Lout$ data points included in the attack training dataset is distorted.
It is not easy to construct a well-trained model from such a complicated structure of data with a linear model such as an SVM.
As a result, the attack results in low accuracy.

The accuracy of the entropy-based approach \TransMIAE{} is below 55\% for any settings.
While the entropy-based approach achieves high accuracy for some applications~\cite{Song_arXiv20}, 
the learning-based approach is more effective than the entropy-based approach in our transfer learning setting.

\subsection{Discussions on our attacks based on the confidence values}
\label{sub:discussion}
In this section, we 
discuss the results of our attacks shown in Section~\ref{sub:attacks:MIA} based on the distributions of confidence values.

\begin{figure*}[t]
\begin{center}
 \begin{minipage}{0.245\hsize}
  \centering
  \includegraphics[width=1.0\hsize]{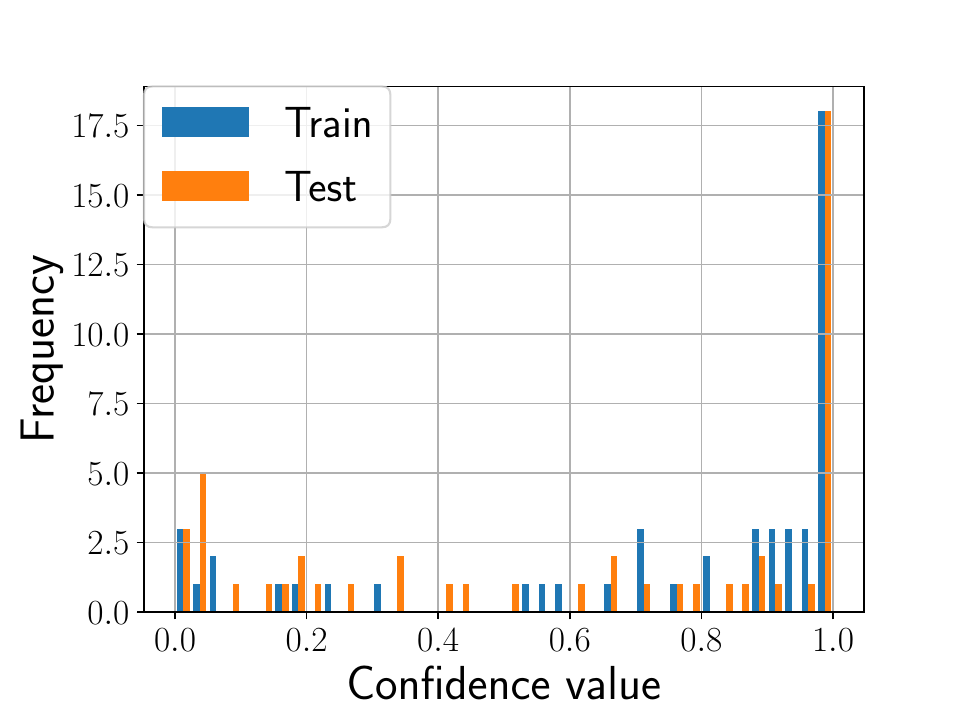}%
  \subcaption{Source model.}
  \label{fig:hist_source_model}
 \end{minipage}
 \begin{minipage}{0.245\hsize}
  \centering
  \includegraphics[width=1.0\hsize]{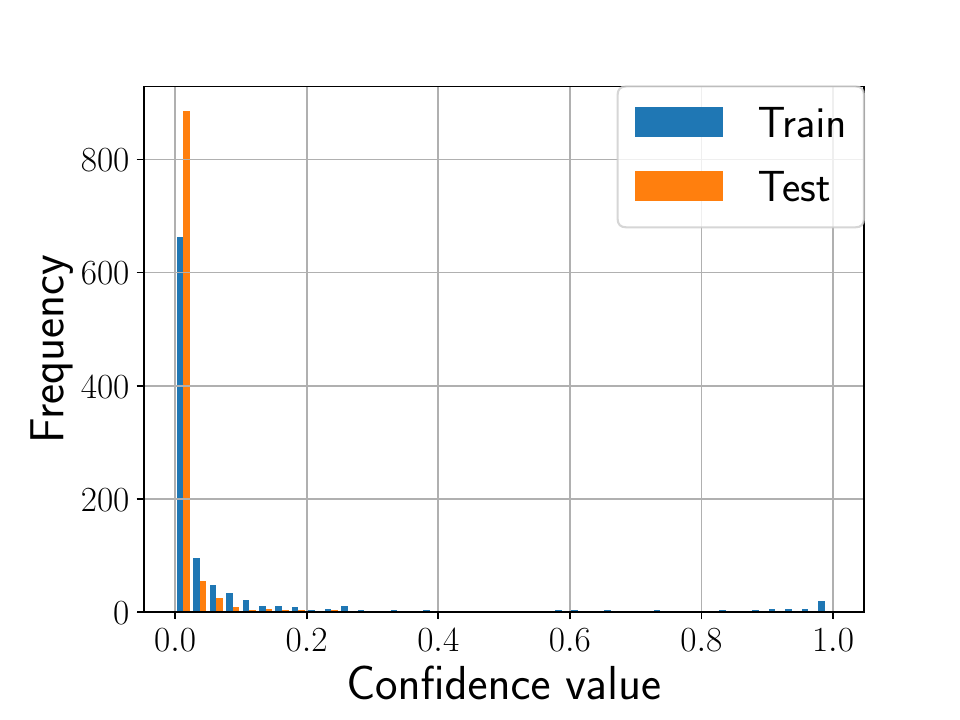}%
  \subcaption{Shadow models (baseline).}
  \label{fig:hist_baseline}
 \end{minipage}
  \begin{minipage}{0.245\hsize}
  \centering
  \includegraphics[width=1.0\hsize]{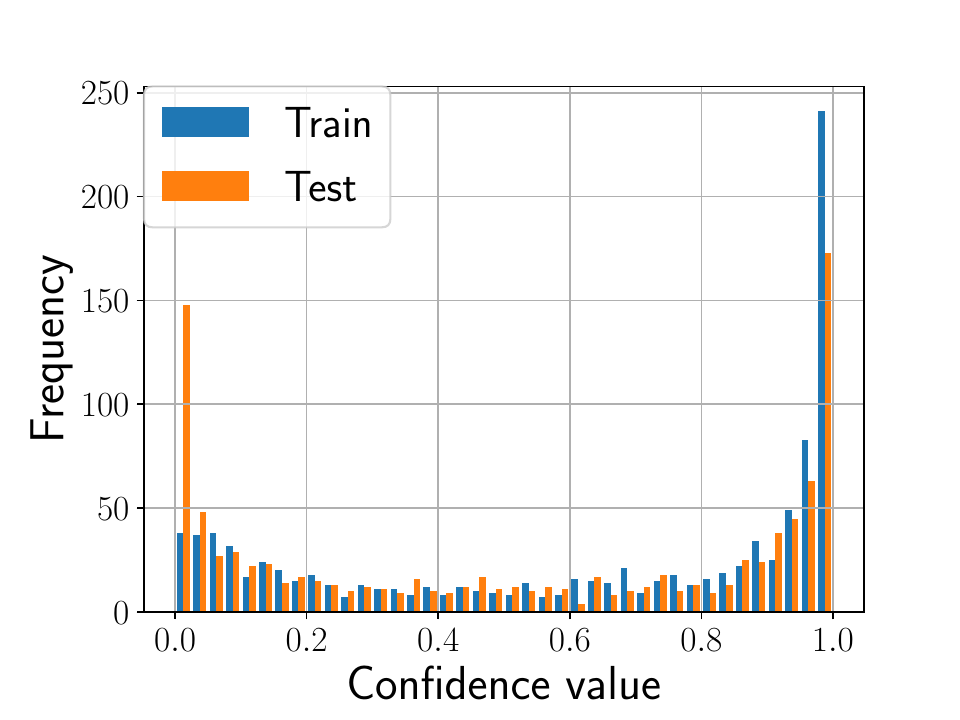}%
  \subcaption{Shadow models (freezing).}
  \label{fig:hist_freezing}
 \end{minipage}
 \begin{minipage}{0.245\hsize}
  \centering
  \includegraphics[width=1.0\hsize]{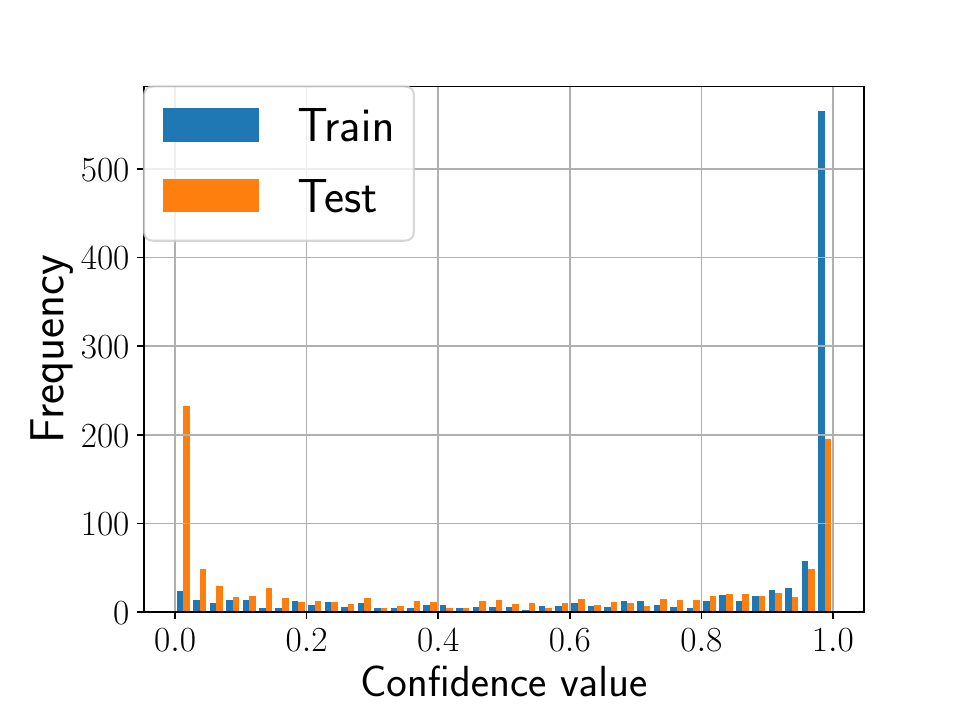}%
  \subcaption{Shadow models (fine-tuning).}
  \label{fig:hist_fine_tuning}
 \end{minipage}
 \caption{Distributions of confidence values of source and shadow models 
 for a single class 
 in the VGGFace2 dataset. Figure~(a) 
 shows 
 the results on the source model. 
 Figures~(b), (c), and (d) 
 show 
 the results on the shadow models constructed with 
 \BaselineMIA{}, 
 \TransMIADNN{} with the freezing approach, and 
 \TransMIADNN{} with fine-tuning approach, 
 respectively. Each graph shows the frequency of confidence values for 
 the training and test data in that class.
 }
 \label{fig:hist_vggface2}
\end{center}
\end{figure*}
\Fig{fig:hist_vggface2} shows the distributions of confidence values of the source model $\fS^i$ and shadow models $\fSH^i$ 
for a single class 
in the VGGFace2 dataset.
The shadow models $\fSH^i$ are constructed with shadow training datasets consisting of 1,000 images.
The graphs~(b), (c), and (d) in \Fig{fig:hist_vggface2} present confidence values for all 100 shadow models.
In these graphs, the shadow models are constructed with \BaselineMIA{}, \TransMIADNN{} with the freezing approach, and \TransMIADNN{} with fine-tuning approach, respectively. 
Each graph consists of both distributions for training and test data.

\smallskip
\noindent{\textbf{Freezing vs fine-tuning.}}~~%
\Fig{fig:hist_vggface2} shows that our transfer shadow training (i.e., the freezing and fine-tuning approaches) trains shadow models 
better than 
the baseline.
The distributions on the freezing approach are the closest to those on the source model.
As a result, the freezing approach achieves the best accuracy among three shadow training approaches.

The distributions 
in 
the fine-tuning approach 
(\Fig{fig:hist_vggface2}-(d)) 
are 
different 
between training and test datasets 
when compared to 
those 
in 
the freezing approach 
(\Fig{fig:hist_vggface2}-(c)). 
More specifically, most of the training data have the confidence value of about 1.0 for the correct label $y$ in the fine-tuning approach (\Fig{fig:hist_vggface2}-(d)).
In other words, 
the shadow models constructed with the fine-tuning approach 
overfit 
shadow training data. 
This explains the results of \Figs{fig:acc_prc_rec} and \ref{fig:frz_fit}, where the freezing approach outperforms the fine-tuning approach in many cases. 

Thus we
conclude that the freezing approach is more appropriate for transfer shadow training than the fine-tuning approach 
when the amount of shadow training data is small.

\smallskip
\noindent{\textbf{Learning-based vs entropy-based.}}~~%
\Fig{fig:hist_vggface2}-(a) 
shows 
that the source model is well trained 
in that 
the distributions for training and test datasets are similar.
In this case, 
the confidence value-based approaches such as MPE (modified prediction entropy) cannot achieve high accuracy, because 
a single threshold does not separate the two distributions.

On the other hand, the learning-based approach 
can deal with localized features of the distribution. 
This is true especially when we use the whole prediction vector as input to the DNN.
Consequently, the 
learning-based 
approach can achieve high accuracy even though the distributions for training and test datasets are similar.
Most of source models in the real-world transfer learning are also well-trained, as large-sale datasets are utilized as the source training dataset. 
This explains the results of Fig.~\ref{fig:frz_fit}, where the learning-based approach outperforms the entropy-based approach.

Note that in \cite{Song_arXiv20}, the MPE (modified prediction entropy)-based membership inference attack outperformed the DNN-based membership inference attack in their experiments. 
When the distributions are very different between training and test datasets (i.e., when the model overfits the training dataset), the MPE-based attack can directly use the difference between the two distributions. 
In that case, the MPE-based attack can outperform the DNN-based attack, because it is not easy to 
find optimal hyperparameters (e.g., hidden layers, learning rate) in the DNN, as pointed out in \cite{Song_arXiv20}.

However, when the distributions are similar between training and test datasets, the entropy-based approach cannot provide high accuracy, as explained above. 
In this case, we should use the learning-based approach.

\subsection{Discussions on Countermeasures}
\label{sub:countermeasures}

We finally discuss possible countermeasures for our transfer learning-based membership inference attacks.

Zou~\etal{}~\cite{Zou_arXiv20} 
considered a privacy attack against transfer learning, and 
showed through experiments that 
an adversary who has black-box access to the target model $\fT$ 
cannot infer membership information about the source training dataset $\DStrain$. 
In \cite{Hidano_arXiv20}, we also show that 
the adversary who has black-box access to the target model $\fT$ cannot infer membership information about $\DStrain$ using the ImageNet dataset. 

We argue that 
both those results, and our results in Section~\ref{sec:evaluation} are important to understand the ``adequate'' level of protection in transfer learning across organizations (described in Section~\ref{sec:intro}). 
Specifically, if a big company (or large hospital) $\cbig$ transfers a part of the source model $\gS$ to a small company (or small hospital) $\cstart$, then the small company $\cstart$ needs to sufficiently protect the transferred model $\gS$ so that the original parameters of $\gS$ are not leaked by $\cstart$. 
Otherwise, the risk of membership inference for $\DStrain$ would be increased by our transfer learning-based membership inference attacks. 

On the other hand, the small company $\cstart$ may allow third parties or end users to have black-box access to the trained target model $\fT$. 
In other words, $\cstart$ may provide a machine learning as a service (MLaaS) 
via an API. 
In this case, the adversary who has black-box access to the target model $\fT$ cannot infer membership information about $\DStrain$, as shown in \cite{Zou_arXiv20,Hidano_arXiv20} 
only by experiments
without theoretical guarantees.

We finally note that the big company $\cbig$ may use \arxiv{regulari-zation-based}\conference{regularization-based} defenses \cite{Nasr:18:CCS,Salem:19:NDSS,Shokri_SP17} or DP (Differential Privacy)-based defenses \cite{Abadi_CCS16,Papernot_ICLR17,Shokri_CCS15,Yu_SP19} to obfuscate the parameters of the transferred model $\gS$ before providing it to $\cstart$. 
If $\gS$'s parameters are sufficiently obfuscated, the membership inference attack can be prevented, even if the obfuscated parameters are leaked via illegal access or internal fraud (just like the local privacy model \cite{Duchi_FOCS13,Kasiviswanathan_FOCS08}). 
However, these approaches have no formal utility-loss guarantees \cite{Jia:19:CCS} and 
can deteriorate the utility when we achieve strong privacy. 
Furthermore, defenses by adding noise to (or transforming) a confidence score vector \cite{Jia:19:CCS,Yang:corr:20:2005-03915} cannot be applied to the whole parameters of $\gS$. 
Obfuscating the parameters of $\gS$ with high utility and privacy is left as future work.

\section{Conclusion}
\label{sec:conclude}
In this paper, we proposed new membership inference attacks 
\TransMIA{} 
where attackers 
use transfer learning to perform membership inference attacks.
In particular, we proposed a \emph{transfer shadow training} technique, which transfers the parameters of the transferred model to a shadow model, to significantly improve the accuracy of membership inference when the amount of shadow training data is small. 
To our knowledge, this work is the first to show that transfer learning can strengthen privacy attacks on machine learning models.

As future work, 
we would like to develop obfuscation methods for the parameters of the transferred model with high utility and privacy, e.g., by extending a game-theoretic approach~\cite{Alvim:17:GameSec}. 
Another line of future work would be to evaluate \TransMIA{} using other types of datasets than image.

\section*{Acknowledgment}
This work was supported by JSPS KAKENHI Grant Number JP19H04113, by ERATO HASUO Metamathematics for Systems Design Project (No. JPMJER1603), JST, and by Inria under the project LOGIS.
In our experiments, we used computational resources of the AI Bridging Cloud Infrastructure (ABCI) provided by National Institute of Advanced Industrial Science and Technology (AIST).

\bibliographystyle{IEEEtran}
\bibliography{main-short}

\appendix
\section{Membership Inference Attacks by Black-box Access to Target Models}
\label{sec:defense}
In this 
appendix, 
we 
evaluate 
membership inference attacks on the transfer learning where an adversary has black-box access to the \emph{target} model $\fT$ and attempts to reveal information on the membership to the \emph{source} training dataset $\DStrain$.

Section~\ref{sub:defense:scenarios} illustrates an attack scenario where an adversary performs membership inference attacks based on black-box access to the target model $\fT$, and
introduces a threat model assumed in this attack. 
Section~\ref{sub:mitigate:transfer} presents the details of the attack algorithms.
Finally, Section~\ref{sub:defense:evaluate} shows experimental results.

\begin{figure}[t]
\centering
\includegraphics[width=0.9\linewidth]{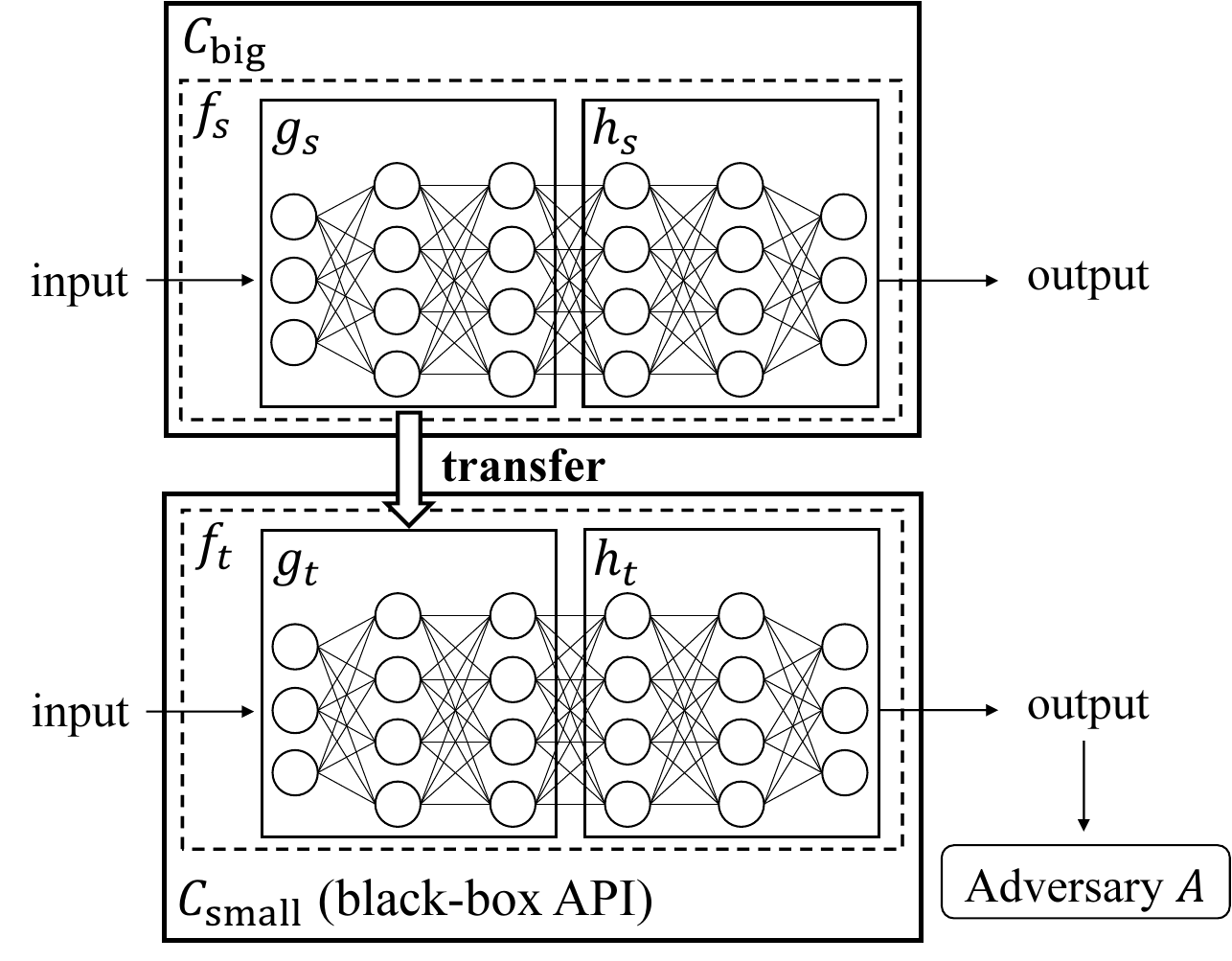}
\caption{Membership inference attack on the source model $\fS$ by black-box access to the target model $\fT$.}
\label{fig:second:scenario}
\end{figure}

\subsection{Attack Scenarios and Threat Model}
\label{sub:defense:scenarios}
We assume another scenario where 
the small company (or small hospital) $\cstart$ offers an API allowing for a black-box access to the target model $\fT$ that they developed by the transfer learning.
Then an adversary $\Atk$, who has black-box access to the target model $\fT$, 
infers 
whether a specific sensitive image $x$ is included in the big company $\cbig$'s source training dataset $\DStrain$ 
(\Fig{fig:second:scenario}).

We investigate a membership inference attack 
that 
aims to determine whether a specific data point $(x, y)$ is included in the source training dataset $\DStrain$.
The 
adversary $\Atk$ 
has 
black-box access to the target network $\fT$ through an API. 
Note that we assume that the adversary $\Atk$ does not obtain the output of the source model $\fS$ upon an arbitrary input (i.e., $\Atk$ does not have black-box access to $\fS$, and have black-box access to only $\fT$), because we want to investigate whether the membership information about $\DStrain$ is leaked by only the output of~$\fT$.

As with the previous work on the membership inference attacks (e.g.,~\cite{Shokri_SP17}), we assume that the adversary $\Atk$ possesses a shadow training dataset $\DSHtrain\subseteq\calx\times\calyS$ and a shadow test dataset $\DSHtest\subseteq\calx\times\calyS$ whose elements were drawn from the same distribution 
$\PS$ as the source training dataset $\DStrain$.
Since the goal of the attack is to reveal information on the membership to the source training dataset $\DStrain$, the adversary $\Atk$ has no prior knowledge of $\DStrain$ itself, hence of whether a data point $(x, y)\in\DSHtrain\cup\DSHtest$ is included in $\DStrain$ or not.
To consider the strongest possible adversary, we assume that $\Atk$ possesses the target training dataset~$\DTtrain$ and the target test dataset~$\DTtest$.
We also assume that $\Atk$ knows whether the small company $\cstart$ constructs a target model $\fT$ with the freezing approach or the fine-tuning approach. 

\subsection{Attack Algorithms}
\label{sub:mitigate:transfer}

Next we present our membership inference attacks against the target model $\fT$.
Here we deal with the learning-based adversary, which is shown to outperform the entropy-based adversary in Section~\ref{sec:evaluation}. 
To construct the attack algorithms, we introduce a shadow training dataset $\DSHItrain \subseteq \DSHtrain$ and a shadow test dataset $\DSHItest \subseteq \DSHtest$ for each shadow model $\fSH^i$.
As with the previous work (e.g.,~\cite{Shokri_SP17}), these shadow training datasets (resp. shadow test datasets) for different shadow models may intersect.

We 
construct a learning-based adversary with black-box access to the target model $\fT$ in the following four steps.
\\\\
\underline{\textbf{\MIAtL{} (learning-based adversary accessing $\fT$)}}
\begin{enumerate}
\item[(TL1)]\, We first construct multiple \emph{shadow source models} $\fSHS^i$ to simulate the behavior of the source model $\fS$ as follows.
For each $i = 1, 2, \ldots , \cSH$, we train a shadow source model $\fSHS^i$ using the shadow training dataset $\DSHItrain$.
Let $\gSHS^i$ be the shallow layers obtained by removing the last fully connected layer from $\fSHS^i$.

\item[(TL2)]\, We next construct multiple \emph{shadow target models} $\fSHT^i$ by transfer learning to simulate the behavior of the target model $\fT$ as follows.
For each $i = 1, 2, \ldots , \cSH$, we build a shadow target model $\fSHT^i$ by initializing its shallow layers $\gSHT^i$ as the network $\gSHS^i$ obtained in (TL1) and then training $\fSHT^i$ with the target training dataset~$\DTtrain$.

\item[(TL3)]\, We then obtain a dataset $\DAtrain$ labeled with membership information ($\Lin$ or $\Lout$) by applying each shadow target model $\fSHT^i$ to 
the shadow training dataset $\DSHtrain$ and the shadow test dataset $\DSHtest$. 
Specifically, we define $\DAtrain$ as the set of the records 
$(x, y, \allowbreak \fSHT^i(x), \Lin)$ for each $(x, y)\in\DSHtrain$ and $(x', y,' \fSHT^i(x'), \allowbreak \Lout)$ for each $(x', y')\in\DSHtest$ 
where $i = 1, 2, \ldots , \cSH$.
We call $\DAtrain$ an \emph{attack training dataset}.

\item[(TL4)]\, 
We finally construct a membership inference adversary $\Atk: \calx\times\calyS\times\Dists\calyT \rightarrow \{\Lin, \Lout\}$
as a classification model (e.g., deep neural network, support vector machine) using the attack training dataset $\DAtrain$ constructed in (TL3). 
\end{enumerate}

Given an input $(x,y)\in\calx\times\calyS$, the learning-based adversary \MIAtL{} obtains $\fT(x)$ by black-box access to the target model $\fT$ and outputs the label $\Atk(x,y, \fT(x))$.

Since the adversary $\Atk$ knows whether $\cstart$ constructs a target model $\fT$ with the freezing approach or the fine-tuning approach (as described in Section~\ref{sub:defense:scenarios}), $\Atk$ uses the same approach as $\cstart$. 
That is, if $\cstart$ uses the freezing (resp.~fine-tuning) approach, then $\Atk$ also uses the freezing (resp.~fine-tuning) approach.
As a membership inference adversary model $\Atk$ in \MIAtL{}, we use a deep neural network (DNN), because it outperformed a support vector machine (SVM) in Section~\ref{sec:evaluation}.

\subsection{Experimental Evaluation}
\label{sub:defense:evaluate}
\begin{table*}[t]
 \centering
 \caption{Accuracy of target models in the CIFAR-100 and Dogs datasets.}
 \begin{tabular}{c|cc|cc|cc|cc}
  \hline
  \multirow{3}{*}{Size} & \multicolumn{4}{c|}{CIFAR-100} & \multicolumn{4}{c}{Dogs} \\ \cline{2-9}
  & \multicolumn{2}{c|}{Freezing} & \multicolumn{2}{c|}{Fine-tuning} & \multicolumn{2}{c|}{Freezing} & \multicolumn{2}{c}{Fine-tuning} \\ \cline{2-9}
  & Train & Test & Train & Test & Train & Test & Train & Test\\ \hline
  100 & 0.810 & 0.080 & 0.500 & 0.070 & 0.360	& 0.040 & 0.770	& 0.050	\\
  200 & 0.635 & 0.105 & 0.610 & 0.095	& 0.505	& 0.055 & 0.765 & 0.055 \\
  300 & 0.473 & 0.107 & 0.707 & 0.090	& 0.507	& 0.057 & 0.407	& 0.053	\\
  400 & 0.545 & 0.115 & 0.703 & 0.105	& 0.525 & 0.065 & 0.765	& 0.078	\\
  500 & 0.450 & 0.118 & 0.602 & 0.110	& 0.392	& 0.062	& 0.646	& 0.074 \\
  600 & 0.555 & 0.133 & 0.850 & 0.128	& 0.183	& 0.057	& 0.280	& 0.058	\\
  800 & 0.334 & 0.146 & 0.914 & 0.159	& 0.276	& 0.075	& 0.330	& 0.075	\\
  1,000 & 0.494 & 0.165 & 0.920 & 0.183	& 0.316	& 0.065	& 0.891	& 0.091 \\
  \hline
  \end{tabular}
  \label{tab:accuarcy_target_models}
\end{table*}

\subsubsection{Datasets and Target Models}
We construct a target model $\fT$ with the source model $\fS$ pre-trained in the experiments of Section~\ref{sec:evaluation}.
We especially reuse the model constructed from the \emph{ImageNet} dataset as $\fS$.

As target training datasets $\DTtrain$, we use two real datasets: \emph{CIFAR-100}~\cite{Krizhevsky2009}
 and \emph{Stanford Dogs} (\emph{Dogs})~\cite{Khosla2011}.
In this section we explain these datasets and the target classification models~$\fT$.

\smallskip
\noindent{\textbf{CIFAR-100.}}~~%
\emph{CIFAR-100} is an image dataset that has 100 classes each consisting of 600 images.
We randomly select the same number of images from each class, and construct 8 target training datasets with different sizes.
We set the size of a target training dataset $\DTtrain$ to 100, 200, 300, 400, 500, 600, 800, or 1,000.
We also construct a target test dataset $\DTtest$ that corresponds to each target training dataset $\DTtrain$ and consists of the same number of images as $\DTtrain$.
We train a target model $\fT$ from each target training dataset with the freezing and fine-tuning approaches.
In the training of each model, we use all the layers of $\fS$ except the last fully connected layer as the transferred source model $\gS$.
We iterate the training over 120 epochs, and select a model with the best accuracy as $\fT$.
Table~\ref{tab:accuarcy_target_models} shows the accuracy of the model $\fT$ for the target training and test datasets.

\smallskip
\noindent{\textbf{Stanford Dogs (Dogs).}}~~%
\emph{Dogs} is a dataset that has 120 breeds of dog images.
We randomly select 100 breeds from all the breeds.
In the same manner as for the CIFAR-100 dataset, we construct 8 target training and test datasets, and then train a target model $\fT$ with each training dataset. 
Table~\ref{tab:accuarcy_target_models} also shows the accuracy of each target training model in this dataset.

\subsubsection{Experimental Setup}
We assume a scenario where an adversary $\Atk$ possesses the target training dataset $\DTtrain$ as described in Section~\ref{sub:defense:scenarios}.
In this scenario, the adversary $\Atk$ trains a shadow source model $\fSHS^i$\footnote{This is the same as the shadow source model $\fSHS^i$  constructed in the experiments of Section~\ref{sec:evaluation}.} to obtain its shallow layers $\gSHS^i$.
Then $\Atk$ constructs a shadow target model $\fSHT^i$ from $\gSHS^i$ by transfer learning using $\DTtrain$ as explained in Section~\ref{sub:mitigate:transfer}.
We apply the same settings to both the target models generated with the CIFAR-100 and Dogs datasets.

In our experiments, we construct 100 shadow target models $\fSHT^i$ for each target model $\fT$
by the \emph{freezing} or \emph{fine-tuning} approach.
Then we produce an attack training dataset $\DAtrain$ as described in Section~\ref{sub:mitigate:transfer}.
For each class label $y$, we construct an attack model, i.e.,  100 attack models in total.

We perform membership inference attacks with the \emph{learning-based approach} 
using a DNN. 
We denote by \MIAtDNN{} 
the attack with the DNN. 
The DNN has three fully-connected hidden layers with 50, 30, and 5 neurons, and is trained with 50 epochs.
We evaluate attack performance with \emph{accuracy}, \emph{precision}, and \emph{recall}.

\subsubsection{Experimental Results}
\begin{figure}[t!]
 \centering
 \begin{minipage}{0.475\hsize}
  \centering
  \includegraphics[width=1.0\hsize]{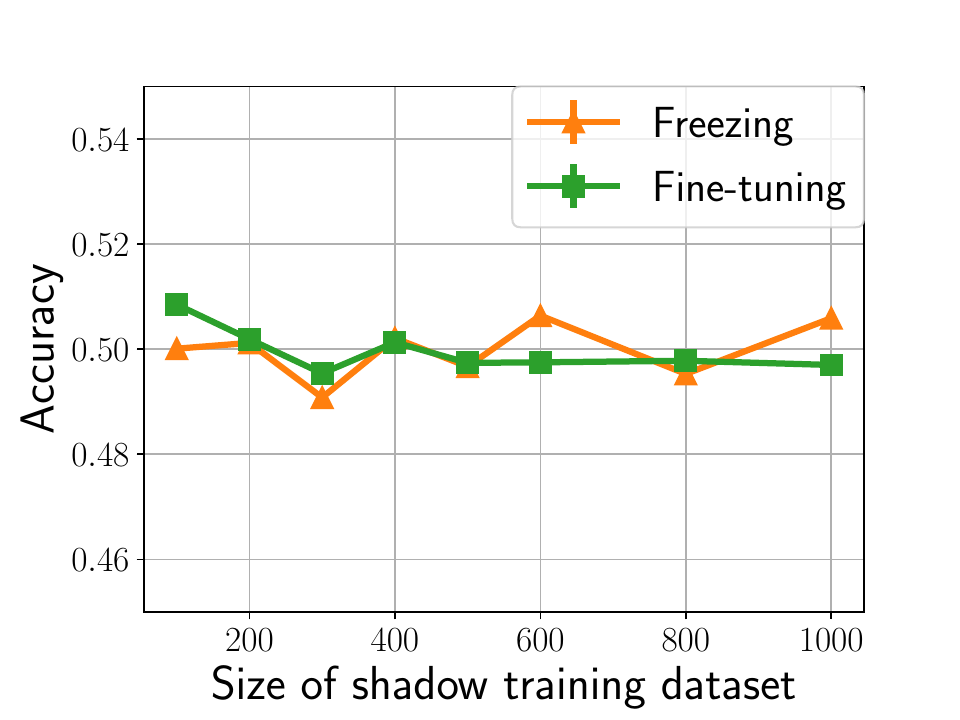}
  \subcaption*{(a) Accuracy in CIFAR-100.}
  \label{fig:cifar100_accuracy}
 \end{minipage}
 \begin{minipage}{0.475\hsize}
  \centering
  \includegraphics[width=1.0\hsize]{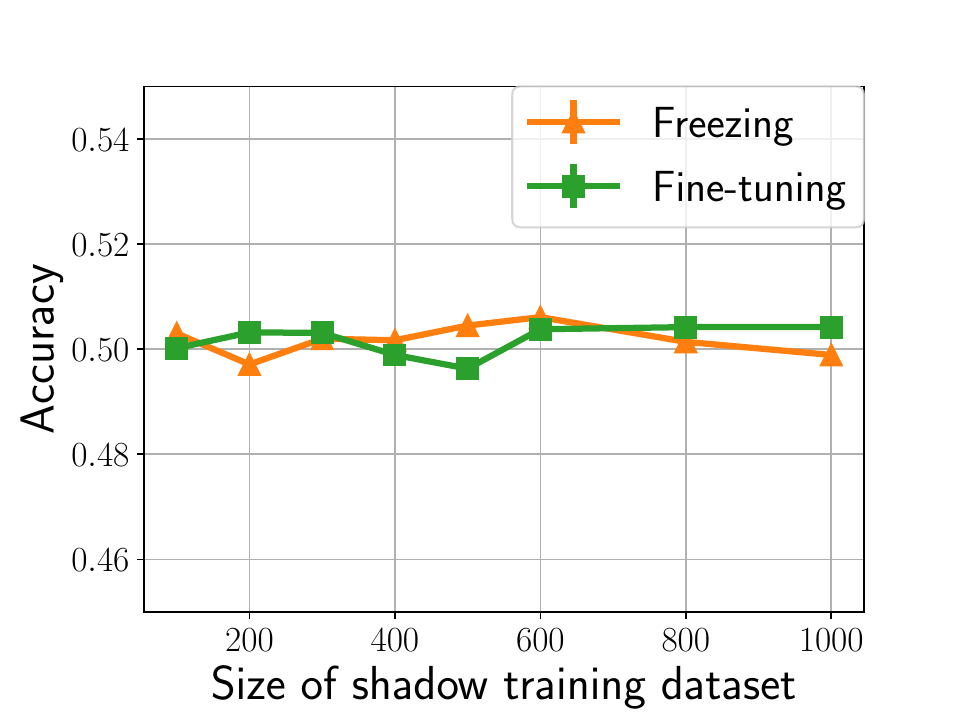}
  \subcaption*{(d) Accuracy in Dogs.}
  \label{fig:dogs_accuracy}
 \end{minipage}
 \begin{minipage}{0.475\hsize}
  \centering
  \includegraphics[width=1.0\hsize]{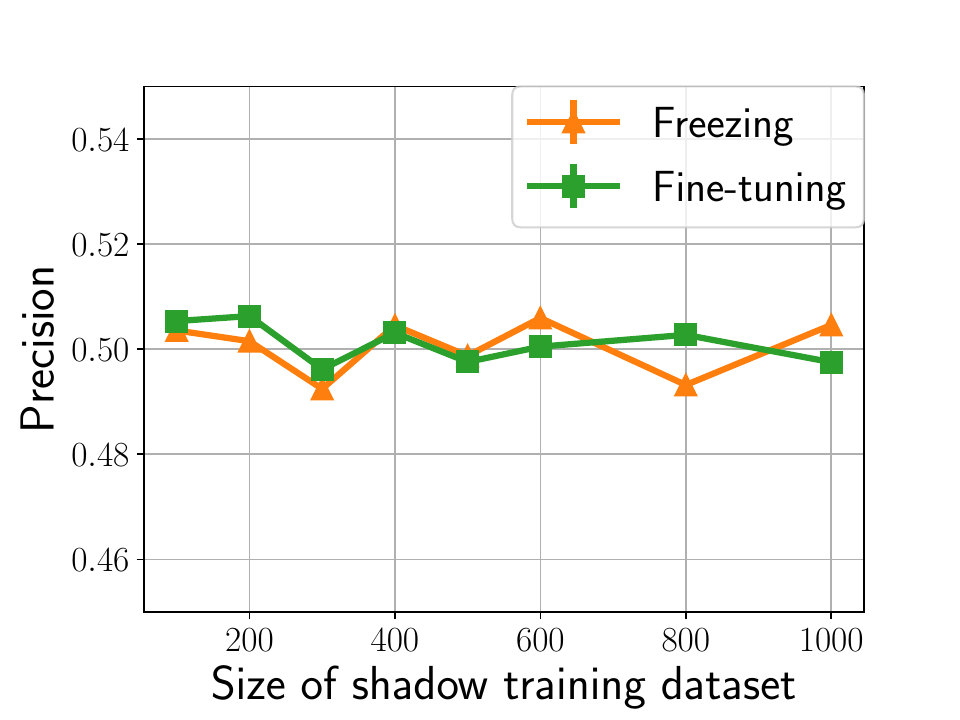}
  \subcaption*{(b) Precision in CIFAR-100.}
  \label{fig:cifar100_precision}
 \end{minipage}
 \begin{minipage}{0.475\hsize}
  \centering
  \includegraphics[width=1.0\hsize]{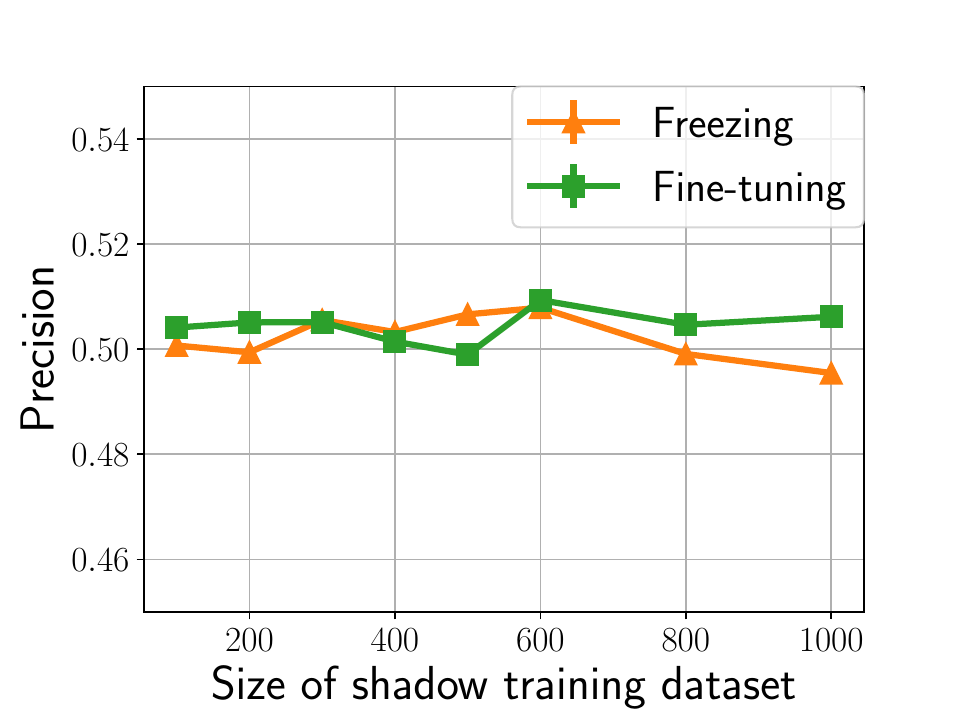}
  \subcaption*{(e) Precision in Dogs.}
  \label{fig:dogs_precision}
 \end{minipage}
 \begin{minipage}{0.475\hsize}
  \centering
  \includegraphics[width=1.0\hsize]{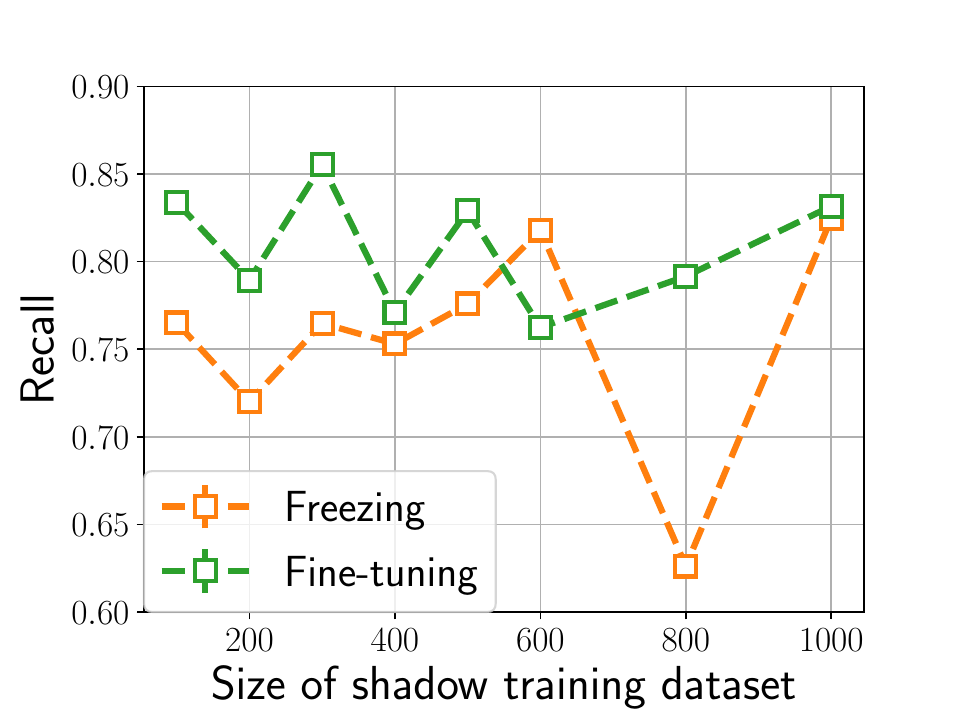}
  \subcaption*{(c) Recall in CIFAR-100.}
  \label{fig:cifar100_recall}
 \end{minipage}
  \begin{minipage}{0.475\hsize}
  \centering
  \includegraphics[width=1.0\hsize]{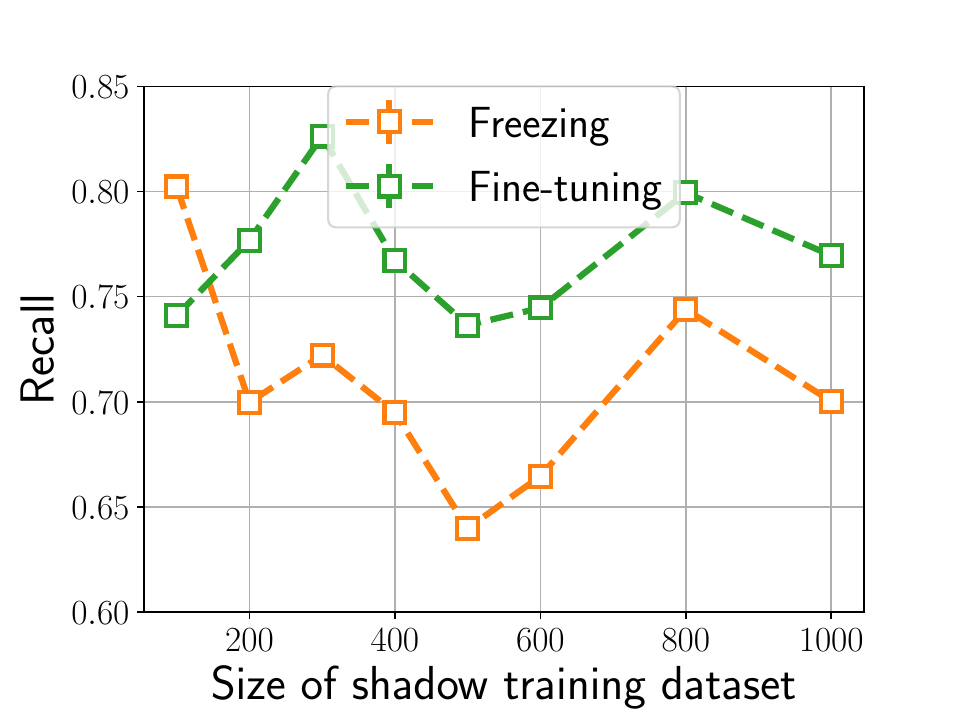}
  \subcaption*{(f) Recall in Dogs.}
  \label{fig:dogs_recall}
 \end{minipage}
 \caption{Accuracy, precision, and recall of 
 our attacks \TransMIADNN{} with the learning-based approach (DNN). 
 In \TransMIADNN{}, we evaluate both 
 freezing and fine-tuning approaches. 
 Figures (a)--(c) (resp. (d)--(f)) are the results in the CIFAR-100 (resp. Dogs) dataset. Each line indicates the average of results for 100 classes. The dashed lines and outline markers in Figures (c) and (f) indicate that the corresponding precision value is less than 60\%. }
 \label{fig:cifar100_dogs_acc_pre_rec}
\end{figure}

We 
present experimental results on the learning-based membership inference attack \MIAtDNN{} with a DNN.
\Fig{fig:cifar100_dogs_acc_pre_rec} shows the accuracy, precision, and recall of the attacks with the freezing and fine-tuning approaches.
Those graphs indicate that the accuracy and precision of \MIAtDNN{} are much lower than those of \TransMIADNN{}, and those values 
are very close to 
0.5.
As mentioned in Section~\ref{sub:attacks:MIA}, when the accuracy and precision are close to 0.5, the adversary fails to infer membership information, even though the corresponding recall is high.

From these results, 
we argue that 
even if the small company $\cstart$ allows the adversary to have black-box access to the trained target model $\fT$, 
the adversary fails to infer membership information about $\DStrain$.
It is left as future work to provide a formal proof that the membership information of the source model $\fS$ is not leaked by the output of the trained target model~$\fT$.

\end{document}